\documentclass[aps,prd,showpacs,onecolumn,preprintnumbers,notitlepage,nofootinbib,tightenlines,10pt]{revtex4-1}
\usepackage{amsmath,bm}
\usepackage{times}
\usepackage{amsfonts}
\usepackage{amssymb}
\usepackage{braket}
\usepackage{color,graphicx}
\usepackage{epstopdf}
\usepackage{latexsym}
\usepackage[dvipsnames]{xcolor}
\usepackage{slashed}
\usepackage{CJK,upgreek,fancyhdr}
\usepackage{float}
\usepackage{multirow}
\usepackage{hyperref}
\hypersetup{colorlinks,citecolor=nicegreen,linkcolor=nicered,urlcolor=RoyalBlue}
\usepackage{enumitem}
\usepackage{natbib}
\usepackage{relsize}
\usepackage[left=2.5cm,right=2.5cm,top=2.0cm,bottom=2.5cm]{geometry}
\newcommand{\beq}{\begin{eqnarray}}
\newcommand{\eeq}{\end{eqnarray}}
\newcommand{\non}{\nonumber\\ }

\def\lsim{ {\ \lower-1.2pt\vbox{\hbox{\rlap{$<$}\lower6pt\vbox{\hbox{$\sim$}
}}}\ } }
\def\gsim{ {\ \lower-1.2pt\vbox{\hbox{\rlap{$>$}\lower6pt\vbox{\hbox{$\sim$}
}}}\ } }

\def \npb{  Nucl. Phys. B }

\def \plb{  Phys. Lett. B }

\def \prd{  Phys. Rev. D }
\def \prl{  Phys. Rev. Lett.  }

\def \jhep{ J. High Energy Phys.  }

\definecolor{Red}{rgb}{1.,0.,0.}
\definecolor{Blue}{rgb}{0.,0.,1.}
\definecolor{RoyalBlue}{rgb}{0.0, 0.14, 0.4}
\definecolor{nicered}{rgb}{0.7,0.1,0.2}
\definecolor{nicegreen}{rgb}{0.1,0.4,0.2}

\def\orcid#1{\kern .08em\href{https://orcid.org/#1}{\includegraphics[keepaspectratio,width=0.76em]{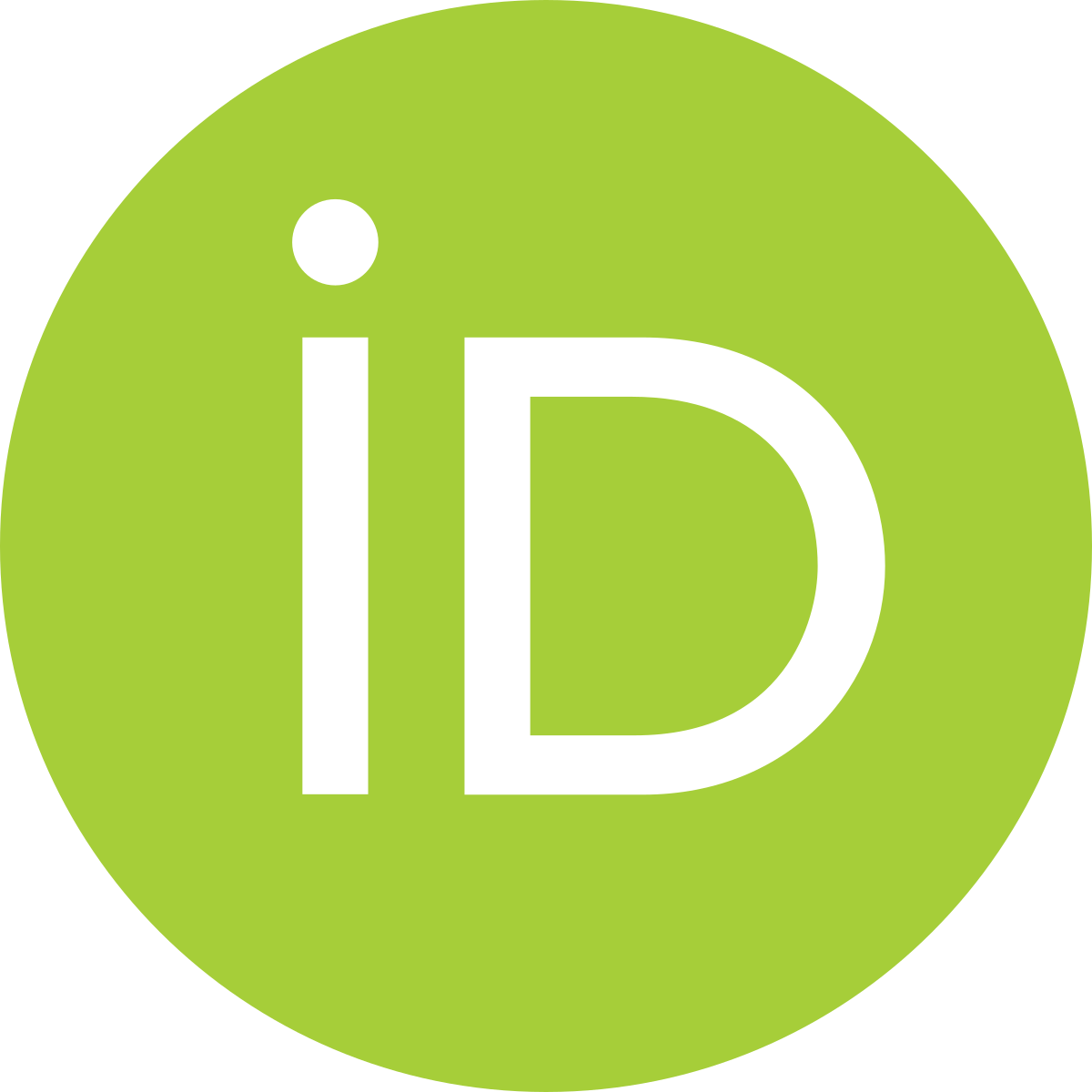}}}

\begin{document}
\title{ Perturbative QCD analysis of neutral \boldmath{$B$}-meson
decays into \boldmath{$\sigma\sigma, \sigma f_0$ and $f_0 f_0$}
}
\author{Hua-Dian~Niu}
\author{Guo-Dong~Li\footnote{\it Li
and Niu contribute equally to this work and are the joint first authors.}}
\affiliation{
C.W.~Chu College,
Jiangsu Normal University, Xuzhou 221116, China}

\author{Jia-Le~Ren}
\email
[ Corresponding author: ]
{renjiale@jsnu.edu.cn}
\author{Xin~Liu\orcid{0000-0001-9419-7462}}
\email
[ Corresponding author: ]
{liuxin@jsnu.edu.cn}
\affiliation{Department of Physics,
Jiangsu Normal University, Xuzhou 221116, China}


\date{\today}

\begin{abstract}

The decays of $B_{d,s}^0 \to \sigma\sigma, \sigma f_0, f_0 f_0$,
with $\sigma$ and $f_0$ denoting the light scalar
mesons $f_0(500)$ and $f_0(980)$ in the two-quark picture, are
studied in the perturbative QCD approach based on $k_T$
factorization. With the referenced value of the mixing angle
$|\varphi| \sim 25^\circ$ for the $\sigma-f_0$ mixing in the quark-flavor basis,
it is of great interest to obtain that:
(a) these neutral $B$-meson decays into $\sigma\sigma, \sigma f_0$, and $f_0 f_0$
have large branching ratios in the order of $10^{-6} \sim 10^{-4}$,
which mean the possibly constructive
interferences existed in the decays with different flavor states, and then are
expected to be tested at the Large Hadron Collider beauty and/or Belle-II
experiments in the (near) future;
(b) the large direct CP violations could be easily found in the
$B_d^0 \to \sigma\sigma, f_0 f_0$ and $B_{d,s}^0 \to \sigma f_0$ decays,
which indicate the considerable interferences between the tree and the
penguin decay amplitudes involved in these four modes, and would be
confronted with the future measurements;
(c) these neutral $B$-meson  decays could be examined through the secondary
decay chain $\sigma/f_0 \to \pi^+ \pi^-$, namely, the four-body
decays of $B_{d,s}^0 \to (\pi^+ \pi^-)_{\sigma(f_0)} (\pi^+ \pi^-)_{\sigma(f_0)}$
with still large branching ratios.
On the other side, it seems that
other 4 four-body decays of $B_d^0 \to (\pi^+ \pi^-)_{\sigma} (K^+ K^-)_{f_0}$,
$B_s^0 \to (\pi^+ \pi^-)_{\sigma(f_0)} (K^+ K^-)_{f_0}$, and $B_s^0 \to (K^+ K^-)_{f_0}
(K^+ K^-)_{f_0}$ could also be detected at the relevant experiments, if the $f_0 \to K^+
K^-$ could be identified from the $\phi \to K^+ K^-$ clearly.
The (near) future experimental confirmations with good precision
would help to further study the perturbative and/or nonperturbative
QCD dynamics involved in these considered decay modes, as well as to
explore the intrinsic characters of these scalar mesons $\sigma$ and/or $f_0$
and to constrain both of the magnitude and the sign of $\varphi$.

\end{abstract}


\preprint{
{\footnotesize JSNU-PHY-HEP-02/21}}
\maketitle


\newpage
%
%
\section{Introduction}\label{sec:intro}

The light scalars have arisen a great of interest at both aspects of
theory and experiment due to the fact that they have the same spin-parity
quantum number, i.e., $J^{P}=0^{+}$, as that of the QCD vacuum. However,
the inner structure of the light scalars is still in controversy theoretically,
though the first observation of the light scalar $f_0(980)$
\footnote{For the sake of simplicity, $f_0$ will be used to denote this
$f_0(980)$ state in the following context, unless otherwise stated.}
state was made by the Belle~\cite{Abe:2002av}
and {\it BABAR}~\cite{Aubert:2003mi} Collaborations through
the decay mode of $B \to f_0 K$. Nevertheless, one is still inspired to explore
the light scalars in the decay products of the heavy $B$ mesons naturally
because of the much larger phase space, with comparison to those produced in the $D_{(s)}$
meson decays. Therefore, investigating on their production in $B$ decays could be
a unique insight to study their underlying structure indeed.

In the conventional quark model, namely, the two-quark picture,
a meson is composed of one quark and one
antiquark, i.e., $q\bar q$, with different coupling of the orbital and
spin angular momenta~\cite{GellMann:1964nj,Zweig:1981pd,Zweig:1964jf}.
Nowadays, the structure of the $S$-wave ground state mesons has almost
been determined unambiguously, though the $\eta$ and $\eta^\prime$
ones contain the possible component of gluonium( or pseudoscalar glueball)
with different extent~\cite{Kou:1999tt,Cheng:2008ss,Liu:2012ib,Li:2021gsx}.
But, the components of the $P$-wave mesons can not be easily determined,
especially of the light scalar states such as $a_0(980)$,
$\kappa$ or $K_0^*(800)$, $\sigma$ or $f_0(500)$, and $f_0(980)$.
Theoretically, many proposals such as $q \bar q$, $\bar q\bar q q q$,
meson-meson bound states, etc. on classifying these light scalars are presented.
It seems that the inner structure is still not well established (for a review,
see e.g., Refs.~\cite{Godfrey:1998pd,Close:2002zu,Tanabashi:2018oca}).
Currently, two different scenarios, namely, Scenario-1(S1) and Scenario-2(S2),
are proposed to classify the light scalars~\cite{Cheng:2005nb}. More specifically, the light
$\sigma$ and $f_0$ are viewed as the $q\bar q$ mesons in S1, while
as the four-quark states in S2. It is necessary to note that the latter structure
is too complicated to be studied in the factorization approach and could not be
quantitatively predicted. Therefore, we will work in S1 for the $\sigma$ and $f_0$
to give several quantitative predictions.

The scalar $\sigma$ and $f_0$ are believed to be made of the superposition of strange and
non-strange quark contents, based on the measurements of $D_s^+ \to f_0 \pi^+$,
$\phi \to f_0 \gamma$, $J/\psi \to f_0 \omega$, $J/\psi \to f_0 \phi$, etc.~\cite{Tanabashi:2018oca}.
For more detail, please see Refs.~\cite{Godfrey:1998pd,Close:2002zu,Tanabashi:2018oca,Amsler:2004ps,Amsler:2004ps,Cheng:2005nb,Klempt:2007cp,Crede:2008vw,Ochs:2013gi}, and references therein. Hence,
analogous to the pseudoscalar $\eta-\eta'$ mixing in the two-quark picture,
the physical $\sigma$ and $f_0$ states can also be described by
a $2\times 2$ rotation matrix with a mixing angle $\varphi$
in the quark-flavor basis, namely,
 \beq
\left(
\begin{array}{c} \sigma \\ f_0 \\ \end{array} \right ) &=&
  \left( \begin{array}{cc}
 \cos{\varphi} & -\sin{\varphi} \\
 \sin{\varphi} & \;\;\;\cos{\varphi} \end{array} \right )
 \left( \begin{array}{c}  f_n\\ f_s \\ \end{array} \right )\;.
 \label{eq:mix-s0-f0}
 \eeq
where $f_n \equiv \frac{u\bar u + d\bar d}{\sqrt{2}}$
and $f_s\equiv s\bar s$ are the quark-flavor states.
It is necessary to mention that the possible scalar glueball
components involved in the $\sigma/f_0$ mesons~\cite{Li:2021gsx}
are left for future investigations elsewhere.
Currently, various measurements on the mixing angle $\varphi$
have been derived and summarized in the
literature with a wide range of values, for example,
see Refs.~\cite{Cheng:2002ai,Cheng:2005nb,Fleischer:2011au,Cheng:2013fba}.
However, an explicit upper limits on the magnitude of the mixing angle
is set as $|\varphi|<31^\circ$ according to the recent Large Hadron Collider
beauty(LHCb) measurements in the $B$ meson decays into $\sigma$ and $f_0$,
i.e., the $B_d^0 \to J/\psi \sigma$ and $B_s^0 \to J/\psi f_0$ decays,
within two-quark structure description~\cite{Aaij:2013zpt}.
On the basis of these data, by also assuming the $\sigma$ and $f_0$ as
$q\bar q$ mesons, a slightly small value of the mixing angle $|\varphi|< 29^\circ$
(90\% C.L.) was proposed by Stone and Zhang~\cite{Stone:2013eaa}.
In Ref.~\cite{Liu:2019ymi},
one of our authors(X.L.) and his collaborators found that the mixing angle $\varphi$
could be further constrained around $25^\circ$ through clarifying the experimental data of
the $B_d^0 \to J/\psi f_0(\to \pi^+ \pi^-)$ and
$B_s^0 \to J/\psi \sigma/f_0(\to \pi^+ \pi^-)$ channels in the two-quark structure for these
two light scalars, except for the challenging $B_d^0 \to J/\psi \sigma(\to \pi^+ \pi^-)$.
Of course, it is unfortunate that the $B_{d,s}^0 \to J/\psi \sigma/f_0$ decays could only
provide the information about the magnitude of $\varphi$ but with a two-fold ambiguity
on the sign. Therefore, as stated in~\cite{Liu:2019ymi}, this ambiguity is expected to be
resolved through the studies of other $B \to M \sigma(f_0)$ decays with $M$ denoting the
open-charmed or light hadrons, once the related measurements are available with high precision.

Presently, several $B \to \sigma/f_0 (P,V)$(Here, $P$ and $V$ denote the pseudoscalar
and vector meson, respectively) decays have been measured experimentally~\cite{Amhis:2016xyh,Tanabashi:2018oca}.
And the related investigations, for instance,
see~\cite{Cheng:2005ye,Cheng:2009xz,Liu:2009xm,Liu:2010zg,Liu:2013cvx,Shen:2006ms,
Wang:2006ria,Wang:2009azc,Colangelo:2010bg,Kim:2009dg,Liu:2021gnp},
are also performed with different approaches/methods theoretically.
With the great development of LHCb and Belle-II experiments~\cite{Belle-II:2018jsg},
more and more modes involving one and/or two scalar states in the $B$ meson
decays are expected to be measured with good precision in the near future.
Therefore, we will systematically study the neutral $B$-meson decays into $\sigma\sigma,
\sigma f_0$, and $f_0 f_0$, i.e., the $B_{d,s}^0 \to \sigma\sigma, \sigma f_0, f_0 f_0$ decays,
by employing the perturbative QCD(PQCD)
approach~\cite{Keum:2000wi,Lu:2000em,Lu:2000hj} based on the $k_T$ factorization theorem.
In Ref.~\cite{Liang:2019eur}, Liang and Yu have analyzed the CP-averaged branching ratios and
CP-violating parameters of the $B_s^0 \to \sigma \sigma$ and
$B_s^0 \to f_0 f_0$ channels in the PQCD approach. However, it is noted that the branching
ratio of the $B_s^0 \to \sigma \sigma$ mode is similar to that of the $B_s^0 \to a_0(980)^0
a_0(980)^0$ one, which seems a bit strange to us that the interferences from the
$B_s^0 \to f_n f_s$ and $f_s f_s$ decay amplitudes did not contribute evidently.
In this work, we will take the possibly considerable interferences arising from the
mixed $f_s$ state with the referenced value of the mixing angle $\varphi \sim 25^\circ$~\cite{Liu:2019ymi}, namely, the contributions from the
$B_s^0 \to f_n f_s$ and $B_s^0 \to
f_s f_s$ decay amplitudes, into account to make reliable predictions in the
$B_{d,s}^0 \to \sigma \sigma, \sigma f_0, f_0 f_0$ channels.

The paper is organized as follows. In Sec.~\ref{sec:form}, we present
the formalism of the PQCD approach and
the related calculations of the considered $B_{d,s}^0 \to \sigma \sigma,
\sigma f_0, f_0 f_0$ decays in a simplified form. Then the numerical calculations and
phenomenological discussions on the related results will be made explicitly
in Sec.~\ref{sec:randd}. The main conclusions and a short summary will be finally
given in Sec.~\ref{sec:summary}.


\section{ Perturbative Calculations}\label{sec:form}

It is well known that the decay amplitude of non-leptonic $B$ meson decays
is determined by the effective and reliable evaluation on the hadronic
matrix element. At the current time, the community provide some standard
factorization approaches/methods, in which the QCD factorization approach~\cite{Beneke99:qcdf,Du:2002up},
the PQCD approach, and the soft-collinear effective
theory~\cite{Bauer:2004tj} are the three more
popular tools based on QCD dynamics presently.
Frankly speaking, due to the existence of the end-point singularities,
the QCD factorization approach and the soft-collinear effective theory
have to parameterize several Feynman amplitudes in the non-factorizable
emission and annihilation diagrams, which finally result in large uncertainties theoretically.
While the PQCD approach, by keeping the transverse momentum($k_T$) of the valence quark,
successfully conquers the end-point singularities that exist in the collinear
factorization theorem. Based on the $k_T$ factorization theorem and armed with the
$k_T$~\cite{Botts89:ktfact,Li92:sudakov}(threshold~\cite{Li02:threshold}) resummation
techniques, the resultant Sudakov factor $S$ (the jet function $J$) could help us
to kill the end-point singularities (smear the double logarithmic divergences). Then
the PQCD approach can be well applied to calculate the hadronic matrix
element of the non-leptonic $B$ meson decays. The decay amplitude could be factorized
into the convolution of the hard kernel associated with the wave functions of the initial
($B$) and final($M_1$ and $M_2$) mesons as follows:
\beq
{\cal A} &\sim& \Phi_B \otimes H \otimes S \otimes J \otimes \Phi_{M_1} \otimes
\Phi_{M_2} \;.
\eeq
where the hard kernel $H$ can be calculated perturbatively in the PQCD approach, while
the wave functions $\Phi$, containing the light-cone distribution amplitudes, are universal
for all modes, although non-perturbative in nature. With the PQCD approach,
the non-factorizable emission diagrams and the annihilation ones can
also be perturbatively calculated, besides the factorizable emission diagrams.
And what's more, the annihilation diagrams perturbatively evaluated in the PQCD approach
can provide a large strong phase that could well explain the CP-violating asymmetries in the
$B$ meson decays~\cite{Chay:2007ep}, for example,
in the $B \to K \pi$ ones~\cite{Keum:2000wi,Keum:2000ph},
which have been confirmed in relevant measurements~\cite{Amhis:2016xyh,Tanabashi:2018oca}
at {\it BABAR}, Belle, and LHCb experiments.
More detail and the recent developments of the PQCD approach could be found
in the literature, for example, see Refs.~\cite{Li:2003yj,Li:2005kt,Li:2012nk,Cheng:2014fwa,Liu:2018kuo,Liu:2020upy,Cheng:2020fcx}.

At the quark level, the $B_{d,s}^0 \to \sigma\sigma, \sigma f_0, f_0 f_0$ decays are
induced by the $\bar b \to \bar d$
or $\bar b \to \bar s$ transitions, respectively.
The weak effective Hamiltonian $H_{\rm eff}$ for
these decays can be written as~\cite{Buchalla:1995vs},
\begin{equation}
H_{\rm eff}\, =\, {G_F\over\sqrt{2}}
\left\{V_{ub}^*V_{uq} \left[C_1(\mu)O_1^{u}(\mu)
+C_2(\mu)O_2^{u}(\mu)\right]- V_{tb}^*V_{tq} \sum_{i=3}^{10}C_i(\mu)O_i(\mu)\right\}\;,
\label{eq:heff}
\end{equation}
with the Fermi constant $G_F=1.16639\times 10^{-5}{\rm GeV}^{-2}$,
the light $q = d, s$ quark, and Wilson
coefficients $C_i(\mu)$ at the renormalization scale
$\mu$. The local four-quark operators $O_i(i=1,\cdots,10)$ are written as
\begin{itemize}
\item{ Tree operators}
\begin{eqnarray}
{\renewcommand\arraystretch{1.5}
\begin{array}{ll}
\displaystyle
O_1^{u}\, =\,
(\bar{q}_\alpha u_\beta)_{V-A}(\bar{u}_\beta b_\alpha)_{V-A}\;,
& \displaystyle
O_2^{u}\, =\, (\bar{q}_\alpha u_\alpha)_{V-A}(\bar{u}_\beta b_\beta)_{V-A}\;;
\end{array}}
\label{eq:operators-1}
\end{eqnarray}

\item{ QCD penguin operators}
\begin{eqnarray}
{\renewcommand\arraystretch{1.5}
\begin{array}{ll}
\displaystyle
O_3\, =\, (\bar{q}_\alpha b_\alpha)_{V-A}\sum_{q'}(\bar{q}'_\beta q'_\beta)_{V-A}\;,
& \displaystyle
O_4\, =\, (\bar{q}_\alpha b_\beta)_{V-A}\sum_{q'}(\bar{q}'_\beta q'_\alpha)_{V-A}\;,
\\
\displaystyle
O_5\, =\, (\bar{q}_\alpha b_\alpha)_{V-A}\sum_{q'}(\bar{q}'_\beta q'_\beta)_{V+A}\;,
& \displaystyle
O_6\, =\, (\bar{q}_\alpha b_\beta)_{V-A}\sum_{q'}(\bar{q}'_\beta q'_\alpha)_{V+A}\;;
\end{array}}
\label{eq:operators-2}
\end{eqnarray}

\item{ Electroweak penguin operators}
\begin{eqnarray}
{\renewcommand\arraystretch{1.5}
\begin{array}{ll}
\displaystyle
O_7\, =\,
\frac{3}{2}(\bar{q}_\alpha b_\alpha)_{V-A}\sum_{q'}e_{q'}(\bar{q}'_\beta q'_\beta)_{V+A}\;,
& \displaystyle
O_8\, =\,
\frac{3}{2}(\bar{q}_\alpha b_\beta)_{V-A}\sum_{q'}e_{q'}(\bar{q}'_\beta q'_\alpha)_{V+A}\;,
\\
\displaystyle
O_9\, =\,
\frac{3}{2}(\bar{q}_\alpha b_\alpha)_{V-A}\sum_{q'}e_{q'}(\bar{q}'_\beta q'_\beta)_{V-A}\;,
& \displaystyle
O_{10}\, =\,
\frac{3}{2}(\bar{q}_\alpha b_\beta)_{V-A}\sum_{q'}e_{q'}(\bar{q}'_\beta q'_\alpha)_{V-A}\;.
\end{array}}
\label{eq:operators-3}
\end{eqnarray}
\end{itemize}
with the color indices $\alpha, \ \beta$ and the notations
$(\bar{q}'q')_{V\pm A} = \bar q' \gamma_\mu (1\pm \gamma_5)q'$.
The index $q'$ in the summation of the above operators runs
through the active quarks $u,\;d,\;s$, $c$, and $b$. It is worth mentioning that
since we work in the framework of the PQCD
approach at leading order[${\cal O}(\alpha_s)$],
it is natural to use the Wilson coefficients at leading order correspondingly.
For the renormalization group evolution of the Wilson coefficients
from higher scale to lower scale, the formulas as given in
Refs.~\cite{Keum:2000wi,Lu:2000em} will be adopted directly.

\begin{figure}[!!htb]
\centering
\begin{tabular}{l}
\includegraphics[width=0.7\textwidth]{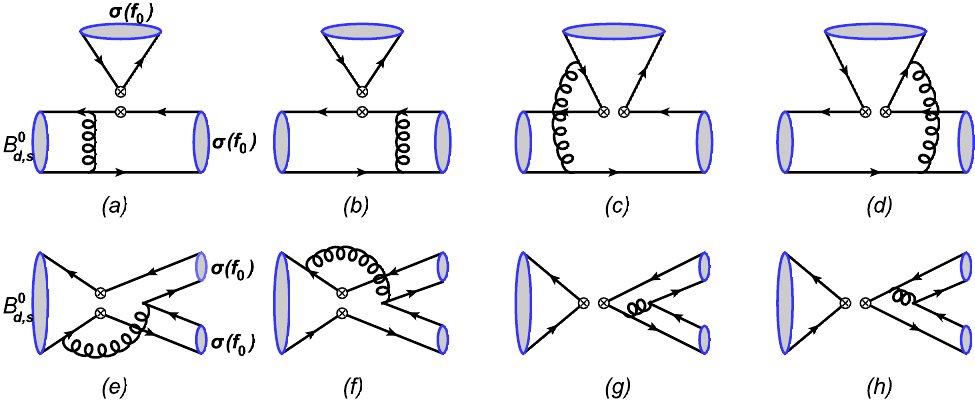}
\end{tabular}
\caption{(Color online) Leading order Feynman diagrams for the
neutral $B$-meson decays into $\sigma\sigma, \sigma f_0$, and $f_0 f_0$
within the framework of PQCD.}
\label{fig:fig1}
\end{figure}

In Fig.~\ref{fig:fig1}, it is clear to see the Feynman diagrams for the
$B_{d,s}^0 \to \sigma\sigma, \sigma f_0, f_0 f_0$ decays in the leading order PQCD
framework and to observe that these diagrams could be classified into two kinds
of topologies, namely, the emission ones with Fig~\ref{fig:fig1}(a)-\ref{fig:fig1}(b)
being the factorizable-emission($fe$) diagrams and Fig~\ref{fig:fig1}(c)-\ref{fig:fig1}(d)
being the non-factorizable-emission($nfe$) diagrams, and the annihilation ones with
Fig~\ref{fig:fig1}(e)-\ref{fig:fig1}(f) being the non-factorizable-annihilation($nfa$)
diagrams and Fig~\ref{fig:fig1}(g)-\ref{fig:fig1}(h) being the factorizable-annihilation($fa$)
diagrams, respectively.
Several two-body non-leptonic $B \to SS$($S$ stands for the scalar meson) decays have
been studied in the PQCD approach by different groups~\cite{Liu:2013lka,Dou:2015mka,Li:2019jlp,Liang:2019eur,Su:2019vbu,Chen:2021dwn},
and the analytic expressions for the factorization formulas and the decay amplitudes
have been presented explicitly in the literature. One just need to identify the scalar meson as the specific
$\sigma$ and/or $f_0$ one to obtain easily the corresponding information of the $B_{d,s}^0 \to
\sigma \sigma, \sigma f_0, f_0 f_0$ decays considered in this work.
Thus, for the sake of simplicity, we do not present the aforementioned formulas
in this paper. The interested readers can refer, for example, to Ref.~\cite{Liu:2013lka}
for detail.

By including the essential contributions arising from the operators as presented in the Eqs.~(\ref{eq:operators-1}), (\ref{eq:operators-2}), and (\ref{eq:operators-3}),
then the decay amplitudes of the considered $B_{d,s}^0 \to\sigma \sigma, \sigma f_0, f_0 f_0$ decays
could be written straightforwardly as follows:
\begin{itemize}

\item[1.]{For $B_d^0 \to\sigma \sigma, \sigma f_0, f_0 f_0$ decays,}

For the sake of convenience, we take the $B_d^0 \to \sigma f_0$ decay as an example to
explain the procedure obtaining the decay amplitudes analytically.
The decay amplitudes for the $B_d^0$ meson decaying into the quark-flavor states
$f_n f_n$, $f_n f_s$, and $f_s f_s$
can be easily written as follows:
\beq
2 A(B_d^0 \to f_n f_n) &=&
V_{ub}^* V_{ud} \biggl\{ a_2 f_{B_d^0} F_{fa}
 + C_2 (M_{nfe} + M_{nfa} ) \biggr\}
- V_{tb}^* V_{td} \biggl\{ ( a_6 -\frac{1}{2} a_8 )
\bar f_{f_n} F^{P_2}_{fe}
\non &&
+ \biggl[2 a_3+a_4+ 2a_5 +\frac{1}{2}(a_7
+ a_{9}- a_{10})\biggr] f_{B_d^0} F_{fa}
+ (a_6 - \frac{1}{2} a_8)f_{B_d^0} F_{fa}^{P_2}
\non &&
+\biggl[C_3 + 2 C_4 - \frac{1}{2} (C_9
-C_{10})\biggr] (M_{nfe} + M_{nfa})
+(C_5 -\frac{1}{2} C_7) (M_{nfe}^{P_1} + M_{nfa}^{P_1})
\non &&
+(2 C_6 + \frac{1}{2} C_8) (M_{nfe}^{P_2}
+ M_{nfa}^{P_2} )\biggr\}\;,
\label{eq:fnfn-d}
\eeq
\beq
\sqrt{2} A(B_d^0 \to f_n f_s)&=&
-V_{tb}^* V_{td} \biggl\{ (C_4 -\frac{1}{2} C_{10})
{M}_{nfe} +(C_6 - \frac{1}{2} C_8)
{M}_{nfe}^{P_2} \biggr\}\;,
\label{eq:fnfs-d}
\eeq
\beq
A(B_d^0 \to f_s f_s)&=&
-V_{tb}^* V_{td} \biggl\{ \biggl[a_3+a_5- \frac{1}{2} (a_7 + a_9)\biggr]
f_{B_d^0} F_{fa}
+(C_4 -\frac{1}{2} C_{10}) M_{nfa}
+(C_6 - \frac{1}{2} C_8)
M_{nfa}^{P_2} \biggr\}\;.
\label{eq:fsfs-d}
\eeq
with $f_{B_d^0}$ and $\bar f_{f_n}$ being the decay constant of initial $B_{d}^0$ meson
and the scalar decay constant of the final flavor state $f_n$.
Notice that the vector decay constants $f_{f_n}$ and $f_{f_s}$ are naturally
zero due to the neutral scalar mesons with the charge conjugation invariance
not being produced by the vector current, which consequently result in the exact
zero factorizable emission contribution $F_{fe}$ in the above decay amplitudes.
It needs to point out that these three equations denote the contributions
with $B_d^0 \to \sigma(f_n)$ transition and here the coefficients, namely,
2, $\sqrt{2}$, and 1,
in front of the amplitudes
$A(B_d^0 \to f_n f_n)$, $A(B_d^0 \to f_n f_s)$, and
$A(B_d^0 \to f_s f_s)$ as correspondingly presented in the Eqs.~(\ref{eq:fnfn-d})-(\ref{eq:fsfs-d}) are from the flavor wave function of $f_n$
and $f_s$. When the $\sigma(f_n)$ and $f_0(f_n)$ states exchange their positions, then these three equations could give the contributions with
$B_d^0 \to f_0(f_n)$ transition in a same form.
All the contributions with both of the $B_d^0 \to \sigma(f_n)$ and the $B_d^0 \to
f_0(f_n)$ transitions will lead to the decay amplitude with physical state, i.e.,
${\cal A}(B_d^0 \to \sigma f_0)$.
In the above formulas, i.e., Eqs.~(\ref{eq:fnfn-d})-(\ref{eq:fsfs-d}),
the ``$F$" and ``$M$" stand for the amplitudes coming from the factorizable
and non-factorizable diagrams associated with the $(V-A)(V-A)$ operators, the $F^{P_1}$ and $M^{P_1}$ stand for the amplitudes coming from the factorizable and non-factorizable
diagrams associated with the $(V-A)(V+A)$ operators, and the $F^{P_2}$ and $M^{P_2}$
stand for the amplitudes coming from the factorizable and non-factorizable diagrams
associated with the $(S-P)(S+P)$ operators that through Fierz transformation of the
$(V-A)(V+A)$ ones, respectively. And the $a_i$ is the standard combination
of the Wilson coefficients $C_i$ defined as follows~\cite{Ali:1998eb}:
\beq
a_1 &=& C_2 + \frac{C_1}{3}\;,\qquad
a_2 = C_1 + \frac{C_2}{3}\;;
\label{eq:Wilson-a-t}
\eeq
\beq
a_i&=& \left\{ \begin{array}{ll}
C_i + C_{i+1}/3 \;\;\;\;\; (i=3,5,7,9),&  \\
C_i + C_{i-1}/3 \;\;\;\;\; (i=4,6,8,10).&   \\ \end{array} \right.
\label{eq:Wilson-a-p}
\eeq
where $C_2 \sim 1$ is
the largest one among all the Wilson coefficients.

By taking the mixing of $\sigma$ and $f_0$ in the quark-flavor basis into account,
the decay amplitude for the physical state of
$B_d^0 \to \sigma f_0$ is then
\beq
{\cal A}(B_d^0 \to \sigma f_0)
&=&\biggl[A(B_d^0 \to f_n f_n)-A(B_d^0 \to f_s f_s)\biggr]\sin(2\varphi)
+A(B_d^0 \to f_n f_s)\cos(2\varphi)\;,
 \label{eq:amp-59-d}
\eeq
Similarly, the $B_d^0 \to \sigma \sigma$ and $B_d^0 \to f_0 f_0$ decay amplitudes
could be written straightforwardly as,
\beq
\sqrt{2} {\cal A}(B_d^0 \to \sigma \sigma)
&=& 2 A(B_d^0 \to f_n f_n)\cos^2\varphi
+2 A(B_d^0 \to f_s f_s)\sin^2\varphi
-A(B_d^0 \to f_n f_s)\sin(2\varphi)\;,
 \label{eq:amp-55-d}
\eeq
\beq
\sqrt{2} {\cal A}(B_d^0 \to f_0 f_0)
&=& 2 A(B_d^0 \to f_n f_n)\sin^2\varphi
+2 A(B_d^0 \to f_s f_s)\cos^2\varphi
 +A(B_d^0 \to f_n f_s)\sin(2\varphi)\;.
 \label{eq:amp-99-d}
\eeq
It is clear to see that these expressions of the decay amplitudes
for the considered modes
are consistent with those for the neutral $B$-meson decays into the
$\eta \eta'$, $\eta \eta$, and $\eta' \eta'$ in the pseudoscalar sector.
For example, please see Ref.~\cite{Ali:2007ff} for detail.

\item[2.]{For $B_s^0 \to \sigma \sigma, \sigma f_0, f_0 f_0$ decays}

Analogously, the decay amplitudes of $B_s^0 \to f_n f_n$, $f_n f_s$,
and $f_s f_s$ can be written as,
\beq
2 A(B_s^0 \to f_n f_n)&=&
V_{ub}^*V_{us}\biggl\{  a_2 f_{B_s^0} F_{fa}
 + C_2 M_{nfa} \biggr\}
- V_{tb}^*V_{ts} \biggl\{(2C_4 + \frac{1}{2} C_{10})
M_{nfa}
\non &&
+ \biggl[2(a_3+a_5)+\frac{1}{2}(a_7+a_9)\biggr] f_{B_s^0} F_{fa}+(2C_6 + \frac{1}{2} C_{8}) M_{nfa}^{P_2}
  \biggr\}\;,
  \label{eq:fnfn-s}
\eeq
\beq
\sqrt{2} A(B_s^0 \to f_n f_s)&=&V_{ub}^*V_{us}
\biggl\{ C_2 M_{nfe} \biggr\}
- V_{tb}^*V_{ts} \biggl\{
(2C_4 + \frac{1}{2} C_{10}) M_{nfe}
+ (2C_6 + \frac{1}{2} C_8) M_{nfe}^{P_2}\biggr\} \;,
\label{eq:fnfs-s}
\eeq
\beq
A(B_s^0 \to f_s f_s)&=& -V_{tb}^*V_{ts}\biggl\{
( a_6 -\frac{1}{2} a_8 )
\bar f_{f_s} F^{P_2}_{fe}
+(C_6 - \frac{1}{2} C_8) (M_{nfe}^{P_2} +M_{nfa}^{P_2} )
\non &&
+\biggl[a_3+a_4+a_5-\frac{1}{2}(a_7+a_9+a_{10}) \biggr]
f_{B_s^0} F_{fa}
\non &&
+(a_6-\frac{1}{2}a_8) f_{B_s^0} F_{fa}^{P_2} +\biggl[C_3 +C_4-\frac{1}{2}(C_9+C_{10})\biggr]
\non && \times
( M_{nfe} +M_{nfa} )
+(C_5 - \frac{1}{2} C_7) (M_{nfe}^{P_1} +M_{nfa}^{P_1} )
\biggr\}\;.
\label{eq:fsfs-s}
\eeq
with $f_{B_s^0}$ and $\bar f_{f_s}$ being the decay constant of the initial
$B_s^0$ meson and the scalar decay constant of the final flavor state $f_s$.
Then, we could give the decay amplitudes for the physical states similarly,
\beq
{\cal A}(B_s^0 \to \sigma f_0)
&=&\biggl[A(B_s^0 \to f_n f_n) - A(B_s^0 \to f_s f_s)\biggr] \sin(2\varphi)+
A(B_s^0 \to f_n f_s) \cos(2\varphi)\;,
\label{eq:amp-59-s}
\eeq
\beq
\sqrt{2} {\cal A}(B_s^0 \to \sigma \sigma)
&=& 2 A(B_s^0 \to f_n f_n)\cos^2\varphi -A(B_s^0 \to f_n f_s) \sin(2\varphi)
+2 A(B_s^0 \to f_s f_s) \sin^2\varphi\;,
\label{eq:amp-55-s}
\eeq
\beq
\sqrt{2} {\cal A}(B_s^0 \to f_0 f_0)
&=& 2 A(B_s^0 \to f_n f_n)\sin^2\varphi +A(B_s^0 \to f_n f_s) \sin(2\varphi)
+2 A(B_s^0 \to f_s f_s) \cos^2\varphi \;.
\label{eq:amp-99-s}
\eeq
\end{itemize}
It is easy to see from the above six decay amplitudes as shown in
Eqs.~(\ref{eq:amp-59-d})-(\ref{eq:amp-99-d}) and
(\ref{eq:amp-59-s})-(\ref{eq:amp-99-s}) that these decay channels
could not only constrain the magnitude but also identify
the sign for the $\sigma$ and $f_0$ mixing angle $\varphi$
by the help of the future measurements with good precision,
due to the possibly significant interferences among the
$B_{d,s}^0 \to f_n f_n$, $B_{d,s}^0 \to f_n f_s$, and
$B_{d,s}^0 \to f_s f_s$ decay amplitudes.


\section{Numerical Results and Discussions} \label{sec:randd}

Now, we come to the numerical calculations of the CP-averaged
branching ratios and the CP-violating asymmetries of the $B_{d,s}^0
\to \sigma \sigma, \sigma f_0, f_0 f_0$ decays in the PQCD approach.
Several comments on the nonperturbative inputs are presented
essentially as follows:
\begin{enumerate}
\item[]{(1)}
For the neutral $B$ mesons, the wave functions (and the distribution
amplitudes) and the decay constants are same as those extensively utilized, for example,
in Refs.~\cite{Keum:2000wi,Lu:2000em,Ali:2007ff,Liu:2013lka}, but with the updated
lifetimes $\tau_{B_d^0} = 1.52$~ps and $\tau_{B_s^0} = 1.509$~ps~\cite{Tanabashi:2018oca}.
The masses of $B_d^0$ and $B_s^0$ mesons are
$m_{B_d^0} = 5.28$~GeV and $m_{B_s^0} = 5.37$~GeV~\cite{Tanabashi:2018oca}, respectively.
The recent developments on the $B$-meson distribution
amplitude could be found in the literature, e.g.,
\cite{Bell:2013tfa,Feldmann:2014ika,Li:2014xda,Braun:2017liq,Wang:2019msf,Galda:2020epp}.
The effects induced by these mentioned
distribution amplitudes could be left for the (near) future investigations with definitely
precise data.

\item[]{(2)}
For the light scalar flavor states, namely, $f_n$ and $f_s$,
the decay constants and the Gegenbauer moments in the distribution
amplitudes have been derived in the QCD sum rule method~\cite{Cheng:2005ye}
and their values at the renormalization scale $\mu =1$ GeV are adopted
same as those in Ref.~\cite{Liu:2019ymi}, specifically,
the scalar decay constants $\bar f_{f_n}  \simeq 0.35$~GeV and
$\bar f_{f_s}  \simeq 0.33$~GeV,
and the Gegenbauer moments $B_1^n = -0.92 \pm 0.08$, $B_3^n = -1.00 \pm 0.05$, and
$B_{1,3}^s \simeq 0.8 B_{1,3}^n$~\cite{Cheng:2005ye}.
Moreover, the masses for the physical states $\sigma$ and $f_0$ and the flavor states
$f_n$ and $f_s$ are same as those utilized in Ref.~\cite{Liu:2019ymi}, i.e.,
$m_\sigma = 0.5$~GeV, $m_{f_0} = 0.98$~GeV, $m_{f_n} = 0.99$~GeV, and
$m_{f_s} = 1.02$~GeV, respectively~\footnote{
Notice that,
for the masses of the $f_n$ and $f_s$ states, the values could also be given
through the mass relation in a different way.
That is, $m^2_{f_n} = m^2_\sigma \cos^2\varphi + m^2_{f_0} \sin^2\varphi$
and $m^2_{f_s} = m^2_{f_0} \cos^2\varphi + m^2_\sigma \sin^2\varphi$~\cite{Cheng:2002ai}.
However, it is worth
stressing that the branching ratios of the considered neutral $B$-meson decays
into $\sigma \sigma$, $\sigma f_0$, and $f_0 f_0$ by employing the masses obtained
with the mass relation are generally consistent with
those by adopting the masses obtained with the QCD sum rule method in this work within
the still large theoretical errors.
}.

\item[]{(3)} For the Cabibbo-Kobayashi-Maskawa(CKM) matrix elements, we also adopt the
Wolfenstein parametrization at leading order,
i.e., up to ${\cal O}(\lambda^4)$,
\beq
V_{\mbox{{\scriptsize CKM}}} &=& \left(\begin{array}{ccc}
1-\frac{1}{2}\lambda^2 \qquad & \lambda \qquad & A\lambda^3(\rho-i\eta) \\
-\lambda \qquad & 1-\frac{1}{2}\lambda^2 \qquad & A\lambda^2\\
A\lambda^3(1-\rho-i\eta) \qquad & -A\lambda^2 \qquad & 1
\end{array}\right)+{\cal O}(\lambda^4)\;,
\label{eq:wolfenstein}
\eeq
but with the updated parameters $A=0.836$, $\lambda=0.22453$, $\bar \rho= 0.122^{+0.018}_{-0.017}$, and $\bar \eta= 0.355^{+0.012}_{-0.011}$~\cite{Tanabashi:2018oca}, in which
$\bar \rho \equiv \rho ( 1- \frac{\lambda^2}{2})$ and $\bar \eta \equiv \eta (1- \frac{\lambda^2}{2})$.
\end{enumerate}

\subsection{CP-averaged Branching Ratios}\label{ssec:br}

Now, we present the numerical results of the $B_{d,s}^0 \to \sigma \sigma, \sigma f_0,
f_0 f_0$ decays in the PQCD approach at leading order.
Firstly, the PQCD predictions of the CP-averaged branching ratios
at the referenced value of the mixing angle $\varphi \sim 25^\circ$
can be read as follows:
\beq
{\cal B}(B_d^0 \to \sigma \sigma) &=&
4.15^{+0.77}_{-0.63}(\omega_B)
^{+1.13}_{-0.95}(B^n_i)
^{+0.05}_{-0.04}(B^s_i)
^{+0.36}_{-0.37}(\varphi)
^{+0.19}_{-0.13}(a_t)
^{+0.21}_{-0.18}(V)
\times 10^{-5}\;,
\label{eq:br-ss-d}\\
{\cal B}(B_d^0 \to \sigma f_0) &=&
2.66^{+0.62}_{-0.50}(\omega_B)
^{+0.61}_{-0.52}(B^n_i)
^{+0.11}_{-0.10}(B^s_i)
^{+0.24}_{-0.26}(\varphi)
^{+0.22}_{-0.16}(a_t)
^{+0.11}_{-0.09}(V)
\times 10^{-5}\;,
\label{eq:br-sf-d}\\
{\cal B}(B_d^0 \to f_0 f_0) &=&
3.36^{+0.54}_{-0.46}(\omega_B)
^{+0.77}_{-0.67}(B^n_i)
^{+0.09}_{-0.09}(B^s_i)
^{+1.26}_{-1.00}(\varphi)
^{+0.36}_{-0.25}(a_t)
^{+0.09}_{-0.08}(V)
\times 10^{-6}\;,
\label{eq:br-ff-d}
\eeq
and
\beq
{\cal B}(B_s^0 \to \sigma \sigma) &=&
1.88^{+0.36}_{-0.29}(\omega_B)
^{+0.33}_{-0.27}(B^n_i)
^{+0.10}_{-0.09}(B^s_i)
^{+0.16}_{-0.16}(\varphi)
^{+0.29}_{-0.23}(a_t)
^{+0.01}_{-0.01}(V)
\times 10^{-4}\;,
\label{eq:br-ss-s}\\
{\cal B}(B_s^0 \to \sigma f_0) &=&
1.22^{+0.00}_{-0.00}(\omega_B)
^{+0.06}_{-0.05}(B^n_i)
^{+0.13}_{-0.12}(B^s_i)
^{+0.04}_{-0.01}(\varphi)
^{+0.37}_{-0.27}(a_t)
^{+0.00}_{-0.00}(V)
\times 10^{-4}\;,
\label{eq:br-sf-s}\\
{\cal B}(B_s^0 \to f_0 f_0) &=&
5.31^{+1.16}_{-0.87}(\omega_B)
^{+0.29}_{-0.28}(B^n_i)
^{+0.62}_{-0.58}(B^s_i)
^{+0.17}_{-0.19}(\varphi)
^{+1.09}_{-0.85}(a_t)
^{+0.01}_{-0.01}(V)
\times 10^{-4}\;.
\label{eq:br-ff-s}
\eeq
It is clearly seen that the CP-averaged branching ratios of the considered
$B_{d,s}^0 \to \sigma \sigma, \sigma f_0, f_0 f_0$ decays vary from
$10^{-6}$ to $10^{-4}$ in the PQCD approach at leading order.
Generally speaking, the largest uncertainties of
these theoretical predictions arise from
the Gegenbauer moments $B_i^n$ and $B_i^s(i=1,3)$, which
are lack of effective constraints at both of the experimental and the theoretical
aspects currently, in the distribution amplitudes of the final
states $f_n$ and $f_s$, as well as from the shape parameter $\omega_B$ in those of the
initial neutral $B$-mesons~\footnote{After all, the only inputs within
the framework of PQCD approach are just the wave-functions ( or distribution amplitudes),
which describe the nonperturbative QCD during the formation of valence quark and valence anti-quark into hadrons. Therefore, the (near) future precise measurements and/or lattice QCD
calculations could be of great importance to constrain these mentioned hadronic inputs.}.
To estimate the possible contributions at higher order, a factor $a_t =1.0 \pm 0.2$ for
the hard scale $t_{\rm max}$, namely, from $0.8t$ to $1.2t$, is introduced to the
numerical calculations, and the resultant results could be considered as one of the sources
of theoretical errors.
The sensitivity of the $B_{d,s}^0 \to \sigma \sigma, \sigma f_0, f_0 f_0$
decay rates to the mixing angle $\varphi$ between the flavor states $f_n$ and $f_s$ is
also presented. The variation of $\varphi$ is taken as 10\% of the central value, namely,
$\varphi =25^\circ \pm 2.5^\circ$, which lead to the relatively smaller theoretical uncertainties
in general, except for that in the $B_d^0 \to f_0 f_0$ mode. Moreover,
it is clear to see that the branching ratios of the $B_d^0$ channels
are more sensitive than those of the $B_s^0$ ones to the variations of the CKM parameters
$V(\bar \rho, \bar \eta)$, which is mainly because both of $|V_{ub}^*V_{ud}|$ and $|V_{tb}^*V_{td}|$ are in the same order, namely, $10^{-3}$, while $|V_{ub}^*V_{us}|$ is
less than $|V_{tb}^*V_{ts}|$ with a factor near 50.
It is noted that two of the considered decays in this work, i.e.,
$B_s^0 \to \sigma \sigma$ and $B_s^0 \to f_0 f_0$, have ever been investigated
in Ref.~\cite{Liang:2019eur}. However, we find that the predicted
values about their branching ratios are a bit smaller than ours in this work,
especially for the $B_s^0 \to \sigma \sigma$ mode
with a highly small decay rate, namely, $4.35^{+1.75}_{-1.50} \times 10^{-6}$.
Certainly,
it is worth mentioning that the Gegenbauer moments $B_1$ and $B_3$
of the flavor states $f_n$ and $f_s$ used
in Ref.~\cite{Liang:2019eur} are different to those adopted in this work,
and the scalar decay constants $\bar f_{f_n}$ and $\bar f_{f_s}$ are slightly
larger than those taken in our evaluations.
We expect the future measurements
at LHCb and/or Belle-II could test these predictions given
by different groups.

In light of these large theoretical errors induced by the hadronic parameters, for the
convenience of future experimental measurements with good precision, several interesting
ratios are defined by employing the above branching ratios in the PQCD approach presented
in the Eqs.~(\ref{eq:br-ss-d})-(\ref{eq:br-ff-d}) and (\ref{eq:br-ss-s})-(\ref{eq:br-ff-s}).
In principle, the uncertainties could be cancelled in the ratios to a great extent, although
the aforementioned hadronic inputs cannot be isolated from the decay amplitudes.
The relevant ratios can be read as follows:
\beq
R_{d\sigma}^{\sigma/f_0}&\equiv&\frac{{\cal B}(B_d^0 \to \sigma \sigma)}
{{\cal B}(B_d^0 \to \sigma f_0)}
= 1.56^{+0.07}_{-0.06}(\omega_B)
^{+0.06}_{-0.07}(B^n_i)
^{+0.05}_{-0.04}(B^s_i)
^{+0.02}_{-0.00}(\varphi)
^{+0.05}_{-0.05}(a_t)
^{+0.01}_{-0.02}(V)\;,
\eeq
\beq
R_{df_0}^{\sigma/f_0}&\equiv&\frac{{\cal B}(B_d^0 \to \sigma f_0)}
{{\cal B}(B_d^0 \to f_0 f_0)}
= 7.92^{+0.49}_{-0.47}(\omega_B)
^{+0.05}_{-0.01}(B^n_i)
^{+0.11}_{-0.09}(B^s_i)
^{+2.25}_{-1.64}(\varphi)
^{+0.12}_{-0.18}(a_t)
^{+0.11}_{-0.08}(V)\;,
\eeq
\beq
R_{d}^{\sigma/f_0}&\equiv&\frac{{\cal B}(B_d^0 \to \sigma \sigma)}
{{\cal B}(B_d^0 \to f_0 f_0)}
= 12.35^{+0.27}_{-0.21}(\omega_B)
^{+0.45}_{-0.47}(B^n_i)
^{+0.22}_{-0.18}(B^s_i)
^{+3.67}_{-2.59}(\varphi)
^{+0.58}_{-0.68}(a_t)
^{+0.29}_{-0.25}(V)\;,
\eeq
\beq
R_{s\sigma}^{\sigma/f_0}&\equiv&\frac{{\cal B}(B_s^0 \to \sigma\sigma)}
{{\cal B}(B_s^0 \to \sigma f_0)}
= 1.54^{+0.30}_{-0.24}(\omega_B)
^{+0.19}_{-0.16}(B^n_i)
^{+0.09}_{-0.08}(B^s_i)
^{+0.08}_{-0.12}(\varphi)
^{+0.20}_{-0.18}(a_t)
^{+0.01}_{-0.01}(V)\;,
\eeq
\beq
R_{sf_0}^{f_0/\sigma}&\equiv&\frac{{\cal B}(B_s^0 \to f_0 f_0)}
{{\cal B}(B_s^0 \to \sigma f_0)}
= 4.35^{+0.95}_{-0.71}(\omega_B)
^{+0.05}_{-0.05}(B^n_i)
^{+0.14}_{-0.16}(B^s_i)
^{+0.00}_{-0.12}(\varphi)
^{+0.34}_{-0.32}(a_t)
^{+0.01}_{-0.01}(V)\;,
\eeq
\beq
R_{s}^{f_0/\sigma}&\equiv&\frac{{\cal B}(B_s^0 \to \ f_0 f_0)}
{{\cal B}(B_s^0 \to \sigma\sigma)}
= 2.82^{+0.07}_{-0.03}(\omega_B)
^{+0.29}_{-0.28}(B^n_i)
^{+0.21}_{-0.20}(B^s_i)
^{+0.16}_{-0.13}(\varphi)
^{+0.13}_{-0.12}(a_t)
^{+0.01}_{-0.01}(V)\;,
\eeq
\beq
R_{s/d}^{\sigma\sigma} &\equiv& \frac{{\cal B}(B_s^0 \to \sigma \sigma)}
{{\cal B}(B_d^0 \to \sigma \sigma)}
=4.53^{+0.02}_{-0.01}(\omega_B)
^{+0.51}_{-0.35}(B^n_i)
^{+0.19}_{-0.17}(B^s_i)
^{+0.02}_{-0.01}(\varphi)
^{+0.47}_{-0.43}(a_t)
^{+0.18}_{-0.20}(V)\;,
\eeq
\beq
R_{s/d}^{\sigma f_0} &\equiv& \frac{{\cal B}(B_s^0 \to \sigma f_0)}
{{\cal B}(B_d^0 \to \sigma f_0)}
=4.59^{+1.06}_{-0.87}(\omega_B)
^{+0.85}_{-0.69}(B^n_i)
^{+0.37}_{-0.36}(B^s_i)
^{+0.45}_{-0.25}(\varphi)
^{+0.93}_{-0.79}(a_t)
^{+0.16}_{-0.19}(V)\;,
\eeq
\beq
R_{s/d}^{f_0 f_0} &\equiv& \frac{{\cal B}(B_s^0 \to f_0 f_0)}
{{\cal B}(B_d^0 \to f_0 f_0)}
=1.58^{+0.08}_{-0.05}(\omega_B)
^{+0.29}_{-0.23}(B^n_i)
^{+0.14}_{-0.14}(B^s_i)
^{+0.59}_{-0.39}(\varphi)
^{+0.14}_{-0.15}(a_t)
^{+0.04}_{-0.04}(V)
\times 10^2\;.
\eeq
One could easily observe that the errors induced by the nonperturbative inputs of the
above ratios are indeed much smaller due to the effective cancellation, except for those of
the ratios $R_{d f_0}^{\sigma/f_0}$, $R_d^{\sigma/f_0}$, and $R_{s/d}^{f_0 f_0}$
because the $B_d^0 \to f_0 f_0$ decay rate is highly sensitive to the mixing angle $\varphi$.
These ratios are expected to be examined in the (near) future experiments at
LHCb and/or Belle-II.

As aforementioned, the neutral $B$-meson decays into $\sigma \sigma, \sigma f_0, f_0 f_0$ would
contain the three decay amplitudes of $B_{d,s}^0 \to f_n f_n$, $B_{d,s}^0 \to f_n f_s$, and
$B_{d,s}^0 \to f_s f_s$ with different ratios when the light scalar $\sigma/f_0$ are treated
as superposition of the $f_n$ and $f_s$ flavor states, which could bring the possibly
constructive or destructive interferences into the $B_{d,s}^0 \to \sigma \sigma, \sigma f_0,
f_0 f_0$ decays.
Within the theoretical uncertainties, the results of the branching ratios for these
neutral $B$-meson decays into $\sigma \sigma, \sigma f_0$, and $f_0 f_0$ by adding
various errors in quadrature could be written explicitly as follows:
\beq
{\cal B}(B_d^0 \to \sigma \sigma)&=& 4.15^{+1.44}_{-1.22} \times 10^{-5}   \;,
\qquad
\hspace{0.22cm}
{\cal B}(B_s^0 \to \sigma \sigma)= 1.88^{+0.60}_{-0.49}  \times 10^{-4}    \;;
\label{eq:br-ss}\\
{\cal B}(B_d^0 \to \sigma f_0) &=& 2.66^{+0.94}_{-0.79}  \times 10^{-5}     \;,
\qquad
\hspace{0.12cm}
{\cal B}(B_s^0 \to \sigma f_0) =  1.22^{+0.40}_{-0.30}  \times 10^{-4}      \;;
\label{eq:br-sf}\\
{\cal B}(B_d^0 \to f_0 f_0) &=&  3.36^{+1.62}_{-1.32}  \times 10^{-6}       \;,
\qquad
{\cal B}(B_s^0 \to f_0 f_0)  =  5.31^{+1.74}_{-1.39}  \times 10^{-4}  \;.
\label{eq:br-ff}
\eeq
Within still large errors, the branching ratios show that ${\cal B}(B_d^0 \to \sigma \sigma)
\sim {\cal B}(B_d^0 \to \sigma f_0) > {\cal B}(B_d^0 \to f_0 f_0)$, and ${\cal B}(B_s^0 \to \sigma \sigma) \sim {\cal B}(B_s^0 \to \sigma f_0) < {\cal B}(B_s^0 \to f_0 f_0)$.
The main reason is
that the $f_n (f_s)$ component dominates the $\sigma (f_0)$ state. In terms of the central
values of the branching ratios, the relation ${\cal B}(B_d^0 \to \sigma \sigma)
> {\cal B}(B_d^0 \to \sigma f_0) > {\cal B}(B_d^0 \to f_0 f_0)$ is easily understood. However,
it is slightly strange that ${\cal B}(B_s^0 \to \sigma \sigma) > {\cal B}(B_s^0 \to
\sigma f_0)$,
which is attributed to the different interferences from the flavorful states $f_n f_n, f_n f_s,$
and $f_s f_s$.

To see the contributions from the diagrams in every topology explicitly,
we present the factorization amplitudes of the considered neutral $B$-meson
decays into $\sigma \sigma, \sigma f_0, f_0 f_0$ in Table~\ref{tab:DecayAmps}, in which
we just quote the central values for clarifications.
The quantities ${\cal A}_{fe}$, ${\cal A}_{nfe}$, ${\cal A}_{nfa}$, and ${\cal A}_{fa}$ are
defined to denote the factorization decay amplitudes arising from the
factorizable emission, the nonfactorizable emission, the nonfactorizable
annihilation, and the factorizable annihilation
diagrams with physical final states, respectively. Specifically, every factorization amplitude includes all the possible
contributions induced by the $(V-A)(V-A)$, $(V-A)(V+A)$, and $(S-P)(S+P)$ currents.
For the sake of the simplicity, in association with the Eqs.~(\ref{eq:fnfn-d})-(\ref{eq:amp-59-d}),
the factorization amplitude ${\cal A}_{fe}$ in the $B_{d,s}^0 \to \sigma f_0$ decay is
taken as an example to clarify the meaning of these four quantities as presented in Table~\ref{tab:DecayAmps} explicitly as follows,
\beq
{\cal A}_{fe}(B_d^0 \to \sigma f_0) &\equiv&
\biggl[ A_{fe}(B_d^0 \to f_n f_n)
- A_{fe}(B_d^0 \to f_s f_s) \biggr]\sin(2\varphi) + A_{fe}(B_d^0 \to f_n f_s) \cos(2\varphi)
\non
&=& A_{fe}(B_d^0 \to f_n f_n) \sin(2\varphi)
\non
&=& \frac{1}{2} \sin(2\varphi)\biggl\{ - V_{tb}^* V_{td} (a_6 -\frac{1}{2}a_8) \bar f_{f_n} F_{fe}^{P_2}  \biggr\}\;,
\eeq
and
\beq
{\cal A}_{fe}(B_s^0 \to \sigma f_0) &\equiv&
\biggl[ A_{fe}(B_s^0 \to f_n f_n)
- A_{fe}(B_s^0 \to f_s f_s) \biggr]\sin(2\varphi) + A_{fe}(B_s^0 \to f_n f_s) \cos(2\varphi)
\non
&=& - A_{fe}(B_s^0 \to f_s f_s) \sin(2\varphi)
\non
&=& \sin(2\varphi)\biggl\{  V_{tb}^* V_{ts} (a_6 -\frac{1}{2}a_8) \bar f_{f_s} F_{fe}^{P_2}  \biggr\}\;.
\eeq
And the other three quantities ${\cal A}_{nfe}$, ${\cal A}_{nfa}$, and ${\cal A}_{fa}$
could also be expressed in a similar manner.
It is interesting to notice that the conventionally large contributions from the
factorizable emission diagrams in the $B \to PP, PV, VV$ decays disappeared
naturally due to the zero vector decay constants $f_{f_n}$ and $f_{f_s}$
in these $B_{d,s}^0 \to \sigma \sigma, \sigma f_0, f_0 f_0$ decays.
In sharp contrast, as stated in Refs.~\cite{Wang:2006ria,Liu:2013lka,Dou:2015mka,Su:2019vbu,Chen:2021dwn},
there are large non-factorizable contributions in the $B$-meson decays into the final states
involving scalar meson(s). In particular, due to the anti-asymmetric leading-twist
distribution amplitude of the scalars, the non-factorizable emission diagrams with a
significant cancellation between Fig.~\ref{fig:fig1}(c) and \ref{fig:fig1}(d) in the pseudoscalar and/or vector sector now become with a dramatic enhancement in the scalar sector, which result
further in the large branching ratios as presented in the Eqs.~(\ref{eq:br-ss})-(\ref{eq:br-ff}).

\begin{table}[htb]
\caption{The factorization decay amplitudes (in units of $10^{-3}$~GeV$^{3}$) of the
$B_{d,s}^0 \to \sigma \sigma, \sigma f_0, f_0 f_0$ decays with the mixing angle $\varphi \sim 25^\circ$
in the PQCD approach at leading order, where only the
central values are quoted for clarifications. }
\label{tab:DecayAmps}
 \begin{center}\vspace{-0.5cm}{
\begin{tabular}[t]{c||c|c|c|c}
\hline  \hline
   Modes   &  ${\cal A}_{fe}$  & ${\cal A}_{nfe}$
   & ${\cal A}_{nfa}$ &  ${\cal A}_{fa}$
    \\
\hline \hline
 $B_d^0 \to \sigma \sigma$
     &$0.873 - {\it i} 0.363$
     &$8.277 + {\it i} 3.283$
     &$-1.663 + {\it i} 1.513$
     &$1.515 + {\it i} 2.007$
   \\
 $B_d^0 \to \sigma f_0$
     &$0.576 - {\it i} 0.240$
     &$7.073 + {\it i} 0.348$
     &$-0.975 + {\it i} 0.269$
     &$0.999 + {\it i} 1.323$
   \\
 $B_d^0 \to f_0 f_0$
     &$0.190 - {\it i} 0.079$
     &$2.864 - {\it i} 0.484$
     &$-0.506 + {\it i} 1.194$
     &$0.329 + {\it i} 0.437$
 \\
\hline \hline
 $B_s^0 \to \sigma \sigma$
     &$-1.341 $
     &$6.905 - {\it i} 5.424$
     &$1.853 - {\it i} 7.537$
     &$-1.030 - {\it i} 4.313$
   \\
 $B_s^0 \to \sigma f_0$
     &$4.066  $
     &$-4.753 + {\it i} 4.268$
     &$0.292 - {\it i} 2.242$
     &$3.138 + {\it i} 13.076$
   \\
 $B_s^0 \to f_0 f_0$
     &$-6.165  $
     &$-17.340 + {\it i} 11.999$
     &$1.506 - {\it i} 4.877$
     &$-4.754 - {\it i} 19.830$
 \\
 \hline \hline
\end{tabular}}
\end{center}
\end{table}

Furthermore, to see clearly the interferences arising from the flavorful states, namely,
$B_{d,s}^0 \to f_n f_n, f_n f_s, f_s f_s$, in these $B_{d,s}^0 \to \sigma \sigma, \sigma
f_0, f_0 f_0$ modes, we also present the decay amplitudes of the neutral $B$-meson decays into
the flavorful and physical final states respectively in Tables~\ref{tab:DecayAmps-FS} and
\ref{tab:DecayAmps-PS}. At the same time, the amplitudes induced by the tree operators and the
penguin operators are also differentiated. From Eqs.~(\ref{eq:fnfn-s})-(\ref{eq:amp-99-s}) and
Table~\ref{tab:DecayAmps-FS}, it is evident to observe that, due to the purely large non-factorizable
emission contributions in the $B_s^0 \to f_n f_s$ decay amplitudes, the slightly constructive
(destructive) interferences between the $B_s^0 \to f_n f_n$ and $B_s^0 \to f_n f_s$ amplitudes consequently
lead to a bit larger (smaller) $B_s^0 \to \sigma \sigma (B_s^0 \to \sigma f_0)$ decay rate.

\begin{table}[htb]
\caption{The decay amplitudes (in units of $10^{-3}$~GeV$^{3}$) of the neutral $B$-meson
decays into the flavorful final states $f_n f_n, f_n f_s$, and $f_s f_s$
in the PQCD approach at leading order, where only the
central values are quoted for clarifications. }
\label{tab:DecayAmps-FS}
 \begin{center}\vspace{-0.5cm}{
\begin{tabular}[t]{c||c|c|c|c}
\hline  \hline
   Flavorful states   & \multicolumn{2}{c|} {${\it A}_{B_d^0}$}
   & \multicolumn{2}{c} {${\it A}_{B_s^0}$}
   \\
   \hline
   & Tree & Penguin & Tree & Penguin
   \\
\hline \hline
 $f_n f_n$
     & \multicolumn{2}{c|} {$17.061 + {\it i} 9.053$}
     & \multicolumn{2}{c} {$2.763 - {\it i} 11.707$}
   \\
   \hline
   & $8.153 + {\it i} 9.593$ & $8.908 - {\it i} 0.540$
   & $-0.566 + {\it i} 0.099$ & $3.329 - {\it i} 11.806$
   \\
\hline \hline
 $f_n f_s$
     & \multicolumn{2}{c|} {$2.283 - {\it i} 2.570$}
     & \multicolumn{2}{c} {$-22.894 + {\it i} 17.226$}
   \\
   \hline
   & $--$ & $2.283 - {\it i} 2.570$
   & $1.922 + {\it i} 1.398$ & $-24.816 + {\it i} 15.828$
   \\
\hline \hline
 $f_s f_s$
     & \multicolumn{2}{c|} {$-0.131 + {\it i} 0.782$}
     & \multicolumn{2}{c} {$-15.783 - {\it i} 15.347$}
   \\
   \hline
   & $--$ & $-0.131 + {\it i} 0.782$
   & $--$ & $-15.783 - {\it i} 15.347$
   \\
\hline \hline
\end{tabular}}
\end{center}
\end{table}

\begin{table}[htb]
\caption{The decay amplitudes (in units of $10^{-3}$~GeV$^{3}$) of the neutral $B$-meson
decays into physical final states $\sigma \sigma, \sigma f_0$, and $f_0 f_0$
in the PQCD approach at leading order, where only the
central values are quoted for clarifications. The mixing angle $\varphi$ is taken as
$25^\circ$.}
\label{tab:DecayAmps-PS}
 \begin{center}\vspace{-0.5cm}{
\begin{tabular}[t]{c||c|c|c|c}
\hline  \hline
   Physical states   & \multicolumn{2}{c|} {${\cal A}_{B_d^0}$}
   & \multicolumn{2}{c} {${\cal A}_{B_s^0}$}
   \\
   \hline
   & Tree & Penguin & Tree & Penguin
   \\
\hline \hline
 $\sigma \sigma$
     & \multicolumn{2}{c|} {$7.673 + {\it i} 1.701$}
     & \multicolumn{2}{c} {$9.033 - {\it i} 24.429$}
   \\
   \hline
   & $3.123 + {\it i} 3.674$ & $4.550 - {\it i} 1.973$
   & $-1.506 - {\it i} 0.676$ & $10.539 - {\it i} 23.753$
   \\
\hline \hline
 $\sigma f_0$
     & \multicolumn{2}{c|} {$12.730 + {\it i} 9.107$}
     & \multicolumn{2}{c} {$2.743 + {\it i} 15.102$}
   \\
   \hline
   & $6.697 + {\it i} 7.879$ & $6.033 + {\it i} 1.228$
   & $0.657 + {\it i} 0.674$ & $2.086 + {\it i} 14.428$
   \\
\hline \hline
 $f_0 f_0$
     & \multicolumn{2}{c|} {$4.070 + {\it i} 1.509$}
     & \multicolumn{2}{c} {$-37.835 - {\it i} 17.971$}
   \\
   \hline
   & $1.456 + {\it i} 1.713$ & $2.613 - {\it i} 0.204$
   & $0.940 + {\it i} 0.775$ & $-38.775 - {\it i} 18.746$
   \\
\hline \hline
\end{tabular}}
\end{center}
\end{table}

It is necessary to point out that the $f_0$ state can decay into $\pi^+ \pi^-$,
as well as into $K^+ K^-$, with the decay rates
\cite{Ablikim:2004cg,Ablikim:2005kp,Aubert:2006nu,Ecklund:2009aa,Aaij:2013zpt}
\beq
{\cal B}(f_0 \to \pi^+ \pi^-)&=& 0.45^{+0.07}_{-0.05}  \;,
\label{eq:br-pp-f0}
\eeq
and
\beq
{\cal B}(f_0 \to K^+ K^-) &=& 0.16^{+0.04}_{-0.05}\;.
\label{eq:br-kk-f0}
\eeq
respectively. Notice that the following assumptions have been made: the decays of
$f_0$ are governed by the $f_0 \to \pi \pi$ and $K K$ modes, and the relations
of the decay rates are $\Gamma(f_0\to \pi^0 \pi^0) = \frac{1}{2} \Gamma(f_0 \to
\pi^+ \pi^-)$ and $\Gamma(f_0 \to K^0 \bar K^0) = \Gamma(f_0 \to K^+ K^-)$.
Moreover, the branching ratio for the $\sigma \to \pi^+ \pi^-$ channel could be
${\cal B}(\sigma \to \pi^+ \pi^-) \simeq 0.67 \pm 0.07$~\cite{Liu:2019ymi}.
Therefore,
one can obtain the following
rich four-body decay channels~\footnote{Very recently, some colleagues began to
study the four-body decays of $B$ mesons in the PQCD approach~\cite{Rui:2021kbn,Li:2021qiw}
with the help of di-meson distribution amplitudes phenomenologically.}
with the possible resonances $\sigma$ and $f_0$
via strong decays into $\pi^+ \pi^-$ in the considered $B_{d,s}^0 \to \sigma \sigma,
\sigma f_0, f_0 f_0$ channels:
\beq
{\rm BR}(B_d^0 \to \sigma(\to \pi^+ \pi^-) \sigma(\to \pi^+ \pi^-) ) &\equiv&
{\cal B}(B_d^0 \to \sigma \sigma) {\cal B}(\sigma \to \pi^+ \pi^-) {\cal B}(\sigma \to \pi^+ \pi^-)
\non &=&
1.86
^{+0.35+0.51+0.02+0.16+0.09+0.09+0.19+0.19}_{-0.28-0.43-0.02-0.17-0.06-0.08-0.19-0.19}
\times 10^{-5}\;,
\eeq
\beq
{\rm BR}(B_d^0 \to \sigma(\to \pi^+ \pi^-) f_0(\to \pi^+ \pi^-)) &\equiv&
{\cal B}(B_d^0 \to \sigma f_0) {\cal B}(\sigma \to \pi^+ \pi^-) {\cal B}(f_0 \to \pi^+ \pi^-)
\non &=&
0.80
^{+0.19+0.18+0.03+0.07+0.07+0.03+0.08+0.12}_{-0.15-0.16-0.03-0.08-0.05-0.03-0.08-0.09}
\times 10^{-5}\;,
\eeq
\beq
{\rm BR}(B_d^0 \to f_0(\to \pi^+ \pi^-) f_0(\to \pi^+ \pi^-)) &\equiv&
{\cal B}(B_d^0 \to f_0 f_0) {\cal B}(f_0 \to \pi^+ \pi^-) {\cal B}(f_0 \to \pi^+ \pi^-)
\non &=&
0.68
^{+0.11+0.16+0.02+0.26+0.07+0.02+0.11+0.11}_{-0.09-0.13-0.02-0.20-0.05-0.02-0.08-0.08}
\times 10^{-6}\;,
\eeq
\beq
{\rm BR}(B_s^0 \to \sigma(\to \pi^+ \pi^-) \sigma(\to \pi^+ \pi^-)) &\equiv&
{\cal B}(B_s^0 \to \sigma \sigma) {\cal B}(\sigma \to \pi^+ \pi^-) {\cal B}(\sigma \to \pi^+ \pi^-)
\non &=&
0.84
^{+0.16+0.15+0.05+0.07+0.13+0.00+0.09+0.09}_{-0.13-0.12-0.04-0.07-0.10-0.00-0.09-0.09}
\times 10^{-4}\;,
\eeq
\beq
{\rm BR}(B_s^0 \to \sigma(\to \pi^+ \pi^-) f_0(\to \pi^+ \pi^-)) &\equiv&
{\cal B}(B_s^0 \to \sigma f_0) {\cal B}(\sigma \to \pi^+ \pi^-) {\cal B}(f_0 \to \pi^+ \pi^-)
\non &=&
0.37
^{+0.00+0.02+0.04+0.01+0.11+0.00+0.04+0.06}_{-0.00-0.02-0.04-0.00-0.08-0.00-0.04-0.04}
\times 10^{-4}\;,
\eeq
\beq
{\rm BR}(B_s^0 \to f_0(\to \pi^+ \pi^-) f_0(\to \pi^+ \pi^-)) &\equiv&
{\cal B}(B_s^0 \to f_0 f_0) {\cal B}(f_0 \to \pi^+ \pi^-) {\cal B}(f_0 \to \pi^+ \pi^-)
\non &=&
1.08
^{+0.23+0.06+0.13+0.03+0.22+0.00+0.17+0.17}_{-0.18-0.06-0.12-0.04-0.17-0.00-0.12-0.12}
\times 10^{-4}\;.
\eeq
in which the last two errors come from the uncertainties of the $\sigma/f_0$ decay width.
All the above modes in association with large numerical results would be explored at the
LHCb and/or Belle-II experiments with good precision in the future.
Moreover, though the $f_0$ resonance coming from the $K^+ K^-$ invariant mass could not be
easily detected at the experimental aspects since the $f_0$ state is usually buried
under the tail of the $\phi$ one, it is essential for us to present the possible
channels induced by the $f_0 \to K^+ K^-$ decay as follows:
\beq
{\rm BR}(B_d^0 \to \sigma(\to \pi^+ \pi^-) f_0(\to K^+ K^-)) &\equiv&
{\cal B}(B_d^0 \to \sigma f_0) {\cal B}(\sigma \to \pi^+ \pi^-) {\cal B}(f_0 \to K^+ K^-)
\non  &=&
0.29
^{+0.07+0.06+0.01+0.03+0.02+0.01+0.03+0.07}_{-0.05-0.05-0.01-0.03-0.02-0.01-0.03-0.09}
\times 10^{-5}\;,
\eeq
\beq
{\rm BR}(B_d^0 \to f_0(\to \pi^+ \pi^-) f_0(\to K^+ K^-)) &\equiv&
{\cal B}(B_d^0 \to f_0 f_0) {\cal B}(f_0 \to \pi^+ \pi^-) {\cal B}(f_0 \to K^+ K^-)
\non &=&
0.24
^{+0.04+0.05+0.01+0.09+0.03+0.01+0.04+0.06}_{-0.03-0.05-0.01-0.07-0.02-0.01-0.03-0.08}
\times 10^{-6}\;,
\eeq
\beq
{\rm BR}(B_d^0 \to f_0(\to K^+ K^-) f_0(\to K^+ K^-)) &\equiv&
{\cal B}(B_d^0 \to f_0 f_0) {\cal B}(f_0 \to K^+ K^-) {\cal B}(f_0 \to K^+ K^-)
\non &=&
0.09
^{+0.01+0.02+0.00+0.03+0.01+0.00+0.02+0.02}_{-0.01-0.02-0.00-0.03-0.01-0.00-0.03-0.03}
\times 10^{-6}\;,
\eeq
\beq
{\rm BR}(B_s^0 \to \sigma(\to \pi^+ \pi^-) f_0(\to K^+ K^-)) &\equiv&
{\cal B}(B_s^0 \to \sigma f_0) {\cal B}(\sigma \to \pi^+ \pi^-) {\cal B}(f_0 \to K^+ K^-)
\non &=&
0.13
^{+0.00+0.01+0.01+0.00+0.04+0.00+0.01+0.03}_{-0.00-0.01-0.01-0.00-0.03-0.00-0.01-0.04}
\times 10^{-4}\;,
\eeq
\beq
{\rm BR}(B_s^0 \to f_0(\to \pi^+ \pi^-) f_0(\to K^+ K^-)) &\equiv&
{\cal B}(B_s^0 \to f_0 f_0) {\cal B}(f_0 \to \pi^+ \pi^-) {\cal B}(f_0 \to K^+ K^-)
\non &=&
0.38
^{+0.08+0.02+0.04+0.01+0.08+0.00+0.06+0.10}_{-0.06-0.02-0.04-0.01-0.06-0.00-0.04-0.12}
\times 10^{-4}\;,
\eeq
\beq
{\rm BR}(B_s^0 \to f_0(\to K^+ K^-) f_0(\to K^+ K^-)) &\equiv&
{\cal B}(B_s^0 \to f_0 f_0) {\cal B}(f_0 \to K^+ K^-) {\cal B}(f_0 \to K^+ K^-)
\non &=&
0.14
^{+0.03+0.01+0.01+0.00+0.03+0.00+0.03+0.03}_{-0.02-0.01-0.01-0.00-0.02-0.00-0.04-0.04}
\times 10^{-4}\;.
\eeq
It is clearly seen that the last three $B_s^0$-meson decay modes have large branching
ratios and are expected to be examined in the near future.

\begin{figure}[!!htb]
\begin{center}
\hspace{-1 cm}
\includegraphics[scale=0.6]{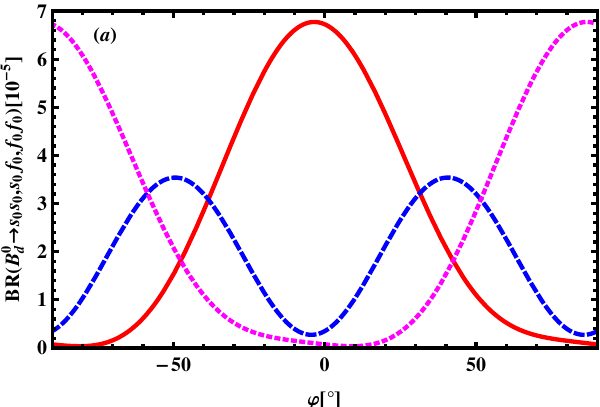}\hspace{1.2cm}
\includegraphics[scale=0.6]{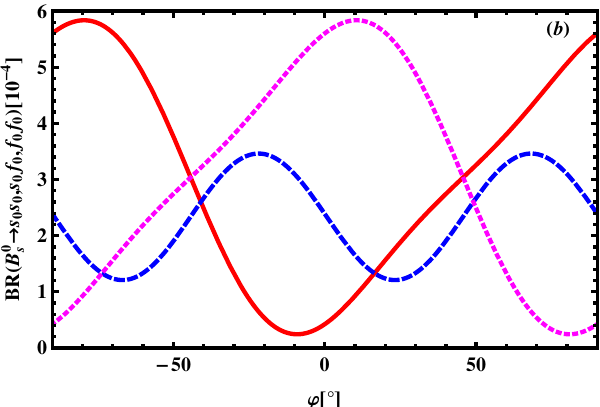}
\caption{(Color online) Dependence of the
CP-averaged $B_{d,s}^0 \to \sigma \sigma, \sigma f_0, f_0 f_0$ branching ratios on
$\varphi$ in the PQCD approach,
in which the red-solid line, the blue-dashed line, and the
magenta-dotted line correspond to the final states of
$\sigma \sigma$,
$\sigma f_0$, and $f_0 f_0$, respectively. }
\label{fig:fig2}
\end{center}
\end{figure}

To provide more information to better constrain the magnitude and the sign
of the mixing angle $\varphi$, we plot the variation of the $B_{d,s}^0 \to
\sigma \sigma, \sigma f_0, f_0 f_0$ decay rates with the mixing angle in the range
of $\varphi \in [-90^\circ, 90^\circ]$ (See Fig.~\ref{fig:fig2}), through which we could find the
differences between the results around $+25^\circ$ and $-25^\circ$, and further
obtain the information about the sign of the mixing angle $\varphi$ once the related
experiments could provide stringent examinations. From Fig.~\ref{fig:fig2}(b),
an interesting variation could be observed that, when the mixing angle are taken as
$-25^\circ$, the relation of the two branching ratios
${\cal B}(B_s^0 \to \sigma \sigma)$ and ${\cal B}(B_s^0 \to \sigma f_0)$ could change from
${\cal B}(B_s^0 \to \sigma \sigma)[1.88 \times 10^{-4}] > {\cal B}(B_s^0 \to \sigma f_0)[1.22
\times 10^{-4}]$ at $\varphi \sim + 25^\circ$
to ${\cal B}(B_s^0 \to \sigma \sigma)[8.75 \times 10^{-5}] < {\cal B}(B_s^0 \to \sigma f_0) [3.43
\times 10^{-4}]$ evidently at $\varphi \sim -25^\circ$.
The underlying reason is that the previously constructive
(destructive) interferences at $\varphi \sim +25^\circ$ become the presently destructive
(constructive) ones at $\varphi \sim -25^\circ$ between the flavorful states
$B_s^0 \to f_n f_n$ and $B_s^0 \to f_n f_s$, which finally result in the considerable
change in the $B_s^0 \to \sigma \sigma$ and $B_s^0 \to \sigma f_0$ decay rates.
Maybe the precise tests in the future on this relation could help us to distinguish
the correct sign of the mixing angle $\varphi$ in the $\sigma-f_0$ mixing.

In order to provide the referenced predictions for the future measurements, even to find the
possible hints for the magnitude and/or sign of $\varphi$, it is essential to present the results
at $\varphi \sim -25^\circ$ for all the above observables that have been shown. Various
predictions in the PQCD approach are presented in order:
\begin{itemize}
\item The CP-averaged branching ratios at $\varphi \sim - 25^\circ$,
\beq
{\cal B}(B_d^0 \to \sigma \sigma) &=&
5.10^{+1.75}_{-1.49}
\times 10^{-5}\;,
\qquad
\hspace{0.22cm}
{\cal B}(B_s^0 \to \sigma \sigma) =
8.75^{+3.20}_{-2.83}
\times 10^{-5}\;;
\label{eq:br-ss-ds-m}
\\
{\cal B}(B_d^0 \to \sigma f_0) &=&
1.69^{+0.67}_{-0.57}
\times 10^{-5}\;,
\qquad
\hspace{0.12cm}
{\cal B}(B_s^0 \to \sigma f_0) =
3.43^{+1.54}_{-1.17}
\times 10^{-4}\;;
\label{eq:br-sf-ds-m}\\
{\cal B}(B_d^0 \to f_0 f_0) &=&
3.61^{+1.14}_{-0.92}
\times 10^{-6}\;,
\qquad
{\cal B}(B_s^0 \to f_0 f_0) =
4.10^{+1.13}_{-0.86}
\times 10^{-4}\;.
\label{eq:br-ff-ds-m}
\eeq

\item Several ratios at $\varphi \sim - 25^\circ$,
\beq
R_{d\sigma}^{\sigma/f_0}&\equiv&\frac{{\cal B}(B_d^0 \to \sigma \sigma)}
{{\cal B}(B_d^0 \to \sigma f_0)}
= 3.02^{+0.36}_{-0.31}
\;,
\qquad
\hspace{0.30cm}
R_{df_0}^{\sigma/f_0} \equiv \frac{{\cal B}(B_d^0 \to \sigma f_0)}
{{\cal B}(B_d^0 \to f_0 f_0)}
= 4.68^{+0.74}_{-0.79}
\;,
\\
R_{d}^{\sigma/f_0}&\equiv&\frac{{\cal B}(B_d^0 \to \sigma \sigma)}
{{\cal B}(B_d^0 \to f_0 f_0)}
= 14.13^{+2.38}_{-2.65}
\;,
\qquad
R_{s\sigma}^{\sigma/f_0}\equiv\frac{{\cal B}(B_s^0 \to \sigma\sigma)}
{{\cal B}(B_s^0 \to \sigma f_0)}
= 0.26^{+0.06}_{-0.07}
\;,
\\
R_{s}^{f_0/\sigma}&\equiv&\frac{{\cal B}(B_s^0 \to \ f_0 f_0)}
{{\cal B}(B_s^0 \to \sigma\sigma)}
= 4.69^{+1.35}_{-1.04}
\;,
\qquad
\hspace{0.10cm}
R_{sf_0}^{f_0/\sigma}\equiv\frac{{\cal B}(B_s^0 \to f_0 f_0)}
{{\cal B}(B_s^0 \to \sigma f_0)}
= 1.20^{+0.42}_{-0.36}
\;,
\\
R_{s/d}^{\sigma\sigma} &\equiv& \frac{{\cal B}(B_s^0 \to \sigma \sigma)}
{{\cal B}(B_d^0 \to \sigma \sigma)}
=1.72^{+0.51}_{-0.43}
\;,
\qquad
\hspace{0.57cm}
R_{s/d}^{\sigma f_0} \equiv \frac{{\cal B}(B_s^0 \to \sigma f_0)}
{{\cal B}(B_d^0 \to \sigma f_0)}
=20.30^{+6.35}_{-5.36}
\;,
\eeq
\beq
R_{s/d}^{f_0 f_0} &\equiv& \frac{{\cal B}(B_s^0 \to f_0 f_0)}
{{\cal B}(B_d^0 \to f_0 f_0)}
=1.14^{+0.31}_{-0.30}
\times 10^2\;.
\eeq

\item Possible four-body decays to $(\pi^+ \pi^-)_{\sigma(f_0)}(\pi^+\pi^-)_{f_0(\sigma)}$ at $\varphi \sim - 25^\circ$,
\beq
{\rm BR}(B_d^0 \to \sigma(\to \pi^+ \pi^-) \sigma(\to \pi^+ \pi^-) )
&=&
2.29^{+0.85}_{-0.75}
\times 10^{-5}\;,\\
{\rm BR}(B_d^0 \to \sigma(\to \pi^+ \pi^-) f_0(\to \pi^+ \pi^-))
&=&
0.51^{+0.22}_{-0.18}
\times 10^{-5}\;,\\
{\rm BR}(B_d^0 \to f_0(\to \pi^+ \pi^-) f_0(\to \pi^+ \pi^-))
&=&
0.73^{+0.16}_{-0.22}
\times 10^{-6}\;,\\
{\rm BR}(B_s^0 \to \sigma(\to \pi^+ \pi^-) \sigma(\to \pi^+ \pi^-))
&=&
3.93^{+1.55}_{-1.40}
\times 10^{-5}\;,\\
{\rm BR}(B_s^0 \to \sigma(\to \pi^+ \pi^-) f_0(\to \pi^+ \pi^-))
&=&
1.03^{+0.51}_{-0.38}
\times 10^{-4}\;,\\
{\rm BR}(B_s^0 \to f_0(\to \pi^+ \pi^-) f_0(\to \pi^+ \pi^-))
&=&
0.83^{+0.29}_{-0.22}
\times 10^{-4}\;.
\eeq

\item Possible four-body decays with $f_0 \to K^+ K^-$ at $\varphi \sim - 25^\circ$,
\beq
{\rm BR}(B_d^0 \to \sigma(\to \pi^+ \pi^-) f_0(\to K^+ K^-))
&=&
0.18^{+0.09}_{-0.09}
\times 10^{-5}\;,\\
{\rm BR}(B_d^0 \to f_0(\to \pi^+ \pi^-) f_0(\to K^+ K^-))
&=&
0.26^{+0.11}_{-0.11}
\times 10^{-6}\;,\\
{\rm BR}(B_d^0 \to f_0(\to K^+ K^-) f_0(\to K^+ K^-))
&=&
0.09^{+0.04}_{-0.05}
\times 10^{-6}\;,\\
{\rm BR}(B_s^0 \to \sigma(\to \pi^+ \pi^-) f_0(\to K^+ K^-))
&=&
0.38^{+0.19}_{-0.17}
\times 10^{-4}\;,\\
{\rm BR}(B_s^0 \to f_0(\to \pi^+ \pi^-) f_0(\to K^+ K^-))
&=&
0.30^{+0.12}_{-0.12}
\times 10^{-4}\;,\\
{\rm BR}(B_s^0 \to f_0(\to K^+ K^-) f_0(\to K^+ K^-))
&=&
0.10^{+0.05}_{-0.05}
\times 10^{-4}\;.
\eeq
\end{itemize}
All the above predictions in the PQCD approach await the future tests with good precision
at LHCb and/or Belle-II, etc.

\subsection{CP-violating Asymmetries}\label{ssec:cp}

Now, let us turn to analyze the CP-violations of the neutral $B$-meson decays into
$\sigma\sigma, \sigma f_0$, and $f_0 f_0$ in the PQCD approach.
In the analysis of the CP-violating asymmetries for the $B_{d,s}^0 \to
\sigma \sigma, \sigma f_0, f_0 f_0$ decays, the effects of neutral $B_{d,s}^0-\bar{B}_{d,s}^0$
mixing should be taken into account. The CP-violating asymmetries
of $B_{d,s}^0(\bar{B}_{d,s}^0) \to \sigma \sigma, \sigma f_0, f_0 f_0$ decays
are time dependent and can be defined as
\beq
A_{\rm CP} &\equiv& \frac{\Gamma\left
(\bar{B}_{d,s}^0(\Delta t) \to f_{\rm CP}\right) -
\Gamma\left(B_{d,s}^0(\Delta t) \to f_{\rm CP}\right )}{ \Gamma\left
(\bar{B}_{d,s}^0(\Delta t) \to f_{\rm CP}\right ) + \Gamma\left
(B_{d,s}^0(\Delta t) \to f_{\rm CP}\right ) }\non
&=& A_{\rm CP}^{\rm dir} \cos(\Delta m_{d,s}  \Delta t)
+ A_{\rm CP}^{\rm mix} \sin (\Delta m_{d,s} \Delta
t)\;, \label{eq:acp-def}
\eeq
where $\Delta m_{d,s}$ is the mass
difference between the two $B_{d,s}^0$ mass eigenstates, $\Delta
t =t_{\rm CP}-t_{tag} $ is the time difference between the tagged
$B_{d,s}^0$ ($\bar{B}_{d,s}^0$) and the accompanying
$\bar{B}_{d,s}^0$ ($B_{d,s}^0$) with opposite $b$ flavor
decaying to the final CP-eigenstate $f_{\rm CP}$ at the time $t_{\rm CP}$.
The direct and mixing-induced CP-violating asymmetries
$A_{\rm CP}^{\rm dir}$ and $A_{\rm CP}^{\rm mix}$
can be written as
\beq
A_{\rm CP}^{\rm dir}&\equiv&
\frac{ \left |
\lambda_{\rm CP}^{d,s}\right |^2-1 } {1+\left |\lambda_{\rm CP}^{d,s}\right |^2},
\qquad
A_{\rm CP}^{\rm mix}\equiv
\frac{ 2 {\rm Im}
(\lambda_{\rm CP}^{d,s})}{1+\left |\lambda_{\rm CP}^{d,s}\right |^2},
\label{eq:acp-csf}
\eeq
where the CP-violating parameter $\lambda_{\rm CP}^{d,s}$ can be read as
\beq
\lambda_{\rm CP}^{d} &\equiv& \eta_f \; \frac{V_{tb}^*V_{td}}{V_{tb}V_{td}^*}
\cdot \frac{ \langle f_{\rm CP} |H_{\rm eff}|\bar{B}_{d}^0\rangle}
{\langle f_{\rm CP} |H_{\rm eff}|B_{d}^0\rangle},
\qquad
\lambda_{\rm CP}^{s} \equiv \eta_f \; \frac{V_{tb}^*V_{ts}}{V_{tb}V_{ts}^*}
\cdot \frac{ \langle f_{\rm CP} |H_{\rm eff}|\bar{B}_{s}^0\rangle}
{\langle f_{\rm CP} |H_{\rm eff}|B_{s}^0\rangle},
\label{eq:lambda-ds}
\eeq
with the CP-eigenvalue of the final states $\eta_f = +1$.
Notice that, for the strange $B$-meson decays, due to the presence of
a non-negligible $\Delta \Gamma_s$, a non-zero ratio $(\Delta
\Gamma/\Gamma)_{B_s^0}$ is expected in the
standard model~\cite{Beneke:1998sy,Fernandez:2006qx}. Thus,
for $B_s^0 \to \sigma \sigma, \sigma f_0, f_0 f_0$ decays, the $\Delta
\Gamma_s$-induced CP-violation
$A_{\rm CP}^{\Delta \Gamma_s}$
can be defined as follows~\cite{Fernandez:2006qx}:
\beq
A_{\rm CP}^{\Delta \Gamma_s} &\equiv& \frac{ 2 {\rm Re}
( \lambda_{\rm CP}^{s})}{1+\left |\lambda_{\rm CP}^{s}\right |^2}.
\label{eq:acp-dgs}
\eeq
The above three quantities describing the CP violations in $B_s^0$
meson decays shown in Eqs.~(\ref{eq:acp-csf}) and
(\ref{eq:acp-dgs}) satisfy the following relation,
\beq
|A_{\rm CP}^{\rm dir}|^2+ |A_{\rm CP}^{\rm mix}|^2
 + |A_{\rm CP}^{\Delta \Gamma_s}|^2 &=&
 1 \;.\label{eq:summation-cp}
\eeq

By the numerical evaluations, the direct and the mixing-induced CP-violating
asymmetries $A_{\rm CP}^{\rm dir}$ and $A_{\rm CP}^{\rm mix}$ for the $B_d^0 \to \sigma \sigma,
\sigma f_0, f_0 f_0$ decays are collected as
\beq
A_{\rm CP}^{\rm dir}(B_d^0 \to \sigma \sigma) &=&
-74.66
^{+4.82}_{-4.58}(\omega_B)
^{+4.57}_{-5.38}(B_i^n)
^{+0.51}_{-0.51}(B_i^s)
^{+1.27}_{-1.33}(\varphi)
^{+4.80}_{-5.32}(a_t)
^{+2.53}_{-2.35}(V)
\times 10^{-2}\;,
\\
A_{\rm CP}^{\rm dir}(B_d^0 \to \sigma f_0) &=&
-37.22
^{+2.89}_{-3.50}(\omega_B)
^{+2.69}_{-3.54}(B_i^n)
^{+1.45}_{-1.39}(B_i^s)
^{+2.49}_{-2.27}(\varphi)
^{+2.89}_{-3.76}(a_t)
^{+1.45}_{-1.46}(V)
\times 10^{-2}\;,
\\
A_{\rm CP}^{\rm dir}(B_d^0 \to f_0 f_0) &=&
-65.96
^{+3.29}_{-2.88}(\omega_B)
^{+3.68}_{-4.18}(B_i^n)
^{+1.31}_{-1.38}(B_i^s)
^{+3.30}_{-4.41}(\varphi)
^{+0.19}_{-0.19}(a_t)
^{+1.85}_{-1.92}(V)
\times 10^{-2}\;;
\eeq
and
\beq
A_{\rm CP}^{\rm mix}(B_d^0 \to \sigma \sigma) &=&
-41.67
^{+3.84}_{-3.43}(\omega_B)
^{+5.50}_{-3.87}(B_i^n)
^{+2.34}_{-2.23}(B_i^s)
^{+3.77}_{-3.44}(\varphi)
^{+1.26}_{-0.09}(a_t)
^{+5.44}_{-5.16}(V)
\times 10^{-2}\;,
\\
A_{\rm CP}^{\rm mix}(B_d^0 \to \sigma f_0) &=&
-92.77
^{+1.56}_{-1.14}(\omega_B)
^{+1.52}_{-0.74}(B_i^n)
^{+0.71}_{-0.58}(B_i^s)
^{+1.24}_{-0.97}(\varphi)
^{+1.78}_{-0.45}(a_t)
^{+0.77}_{-0.58}(V)
\times 10^{-2}\;,
\\
A_{\rm CP}^{\rm mix}(B_d^0 \to f_0 f_0) &=&
-74.31
^{+3.38}_{-3.28}(\omega_B)
^{+5.11}_{-3.68}(B_i^n)
^{+1.60}_{-1.45}(B_i^s)
^{+4.86}_{-3.20}(\varphi)
^{+2.00}_{-0.64}(a_t)
^{+1.61}_{-1.54}(V)
\times 10^{-2}\;;
\eeq
The large direct CP-violating asymmetries indicate that these $B_d^0 \to \sigma \sigma,
\sigma f_0, f_0 f_0$ decays contain the large tree amplitudes and the large penguin
amplitudes simultaneously, which could be evidently seen from the decay amplitudes
as shown in Table~\ref{tab:DecayAmps-PS} and then lead to significant interferences
between these two amplitudes. In light of the predicted large decay rates around
$10^{-6} \sim 10^{-5}$ in the PQCD approach, it is expected that these large direct
CP-violating asymmetries in the considered
neutral $B_d^0$-meson decays into $\sigma \sigma$, $\sigma f_0$, and $f_0 f_0$
could be confronted with the relevant experiments in the future.

And the direct, the mixing, and the $\Delta\Gamma_s$-induced CP violations
$A_{\rm CP}^{\rm dir}$, $A_{\rm CP}^{\rm mix}$, and $A_{\rm CP}^{\rm \Delta\Gamma_s}$
for the $B_s^0 \to \sigma \sigma, \sigma f_0, f_0 f_0$ decays
predicted in the PQCD approach are as follows:
\beq
A_{\rm CP}^{\rm dir}(B_s^0 \to \sigma \sigma) &=&\hspace{0.16cm}
-4.10
^{+0.07}_{-0.01}(\omega_B)
^{+0.28}_{-0.22}(B_i^n)
^{+0.19}_{-0.17}(B_i^s)
^{+0.54}_{-0.54}(\varphi)
^{+0.56}_{-0.46}(a_t)
^{+0.13}_{-0.13}(V)
\times 10^{-2}\;,\\
A_{\rm CP}^{\rm dir}(B_s^0 \to \sigma f_0) &=&
-11.76
^{+1.46}_{-1.65}(\omega_B)
^{+1.26}_{-1.25}(B_i^n)
^{+0.17}_{-0.16}(B_i^s)
^{+1.34}_{-0.62}(\varphi)
^{+1.64}_{-1.71}(a_t)
^{+0.36}_{-0.40}(V)
\times 10^{-2}\;,
\\
A_{\rm CP}^{\rm dir}(B_s^0 \to f_0 f_0) &=&\hspace{0.44cm}
4.55
^{+0.09}_{-0.19}(\omega_B)
^{+0.15}_{-0.17}(B_i^n)
^{+0.10}_{-0.10}(B_i^s)
^{+0.27}_{-0.31}(\varphi)
^{+0.30}_{-0.28}(a_t)
^{+0.15}_{-0.15}(V)
\times 10^{-2}\;.
\eeq
and
\beq
A_{\rm CP}^{\rm mix}(B_s^0 \to \sigma \sigma) &=&
11.50
^{+0.65}_{-0.61}(\omega_B)
^{+0.51}_{-0.61}(B_i^n)
^{+0.08}_{-0.08}(B_i^s)
^{+0.09}_{-0.17}(\varphi)
^{+0.79}_{-0.74}(a_t)
^{+0.38}_{-0.36}(V)
\times 10^{-2}\;,
\\
A_{\rm CP}^{\rm mix}(B_s^0 \to \sigma f_0) &=&\hspace{0.16cm}
3.36
^{+0.06}_{-0.25}(\omega_B)
^{+0.61}_{-0.47}(B_i^n)
^{+0.18}_{-0.17}(B_i^s)
^{+2.71}_{-2.37}(\varphi)
^{+0.22}_{-0.07}(a_t)
^{+0.11}_{-0.12}(V)
\times 10^{-2}\;,
\\
A_{\rm CP}^{\rm mix}(B_s^0 \to f_0 f_0) &=&\hspace{0.16cm}
2.87
^{+0.53}_{-0.58}(\omega_B)
^{+0.42}_{-0.45}(B_i^n)
^{+0.03}_{-0.05}(B_i^s)
^{+0.38}_{-0.38}(\varphi)
^{+0.50}_{-0.52}(a_t)
^{+0.09}_{-0.09}(V)
\times 10^{-2}\;.
\eeq
and
\beq
A_{\rm CP}^{\rm \Delta\Gamma_s}(B_s^0 \to \sigma \sigma) &=&
99.25
^{+0.07}_{-0.07}(\omega_B)
^{+0.07}_{-0.05}(B_i^n)
^{+0.02}_{-0.01}(B_i^s)
^{+0.01}_{-0.00}(\varphi)
^{+0.11}_{-0.11}(a_t)
^{+0.05}_{-0.05}(V)
\times 10^{-2}\;,
\\
A_{\rm CP}^{\rm \Delta\Gamma_s}(B_s^0 \to \sigma f_0) &=&
99.25
^{+0.16}_{-0.20}(\omega_B)
^{+0.15}_{-0.18}(B_i^n)
^{+0.01}_{-0.01}(B_i^s)
^{+0.20}_{-0.21}(\varphi)
^{+0.18}_{-0.23}(a_t)
^{+0.04}_{-0.05}(V)
\times 10^{-2}\;,
\\
A_{\rm CP}^{\rm \Delta\Gamma_s}(B_s^0 \to f_0 f_0) &=&
99.86
^{+0.01}_{-0.01}(\omega_B)
^{+0.01}_{-0.02}(B_i^n)
^{+0.00}_{-0.01}(B_i^s)
^{+0.02}_{-0.03}(\varphi)
^{+0.02}_{-0.03}(a_t)
^{+0.00}_{-0.01}(V)
\times 10^{-2}\;.
\eeq
Different from the $B_d^0 \to \sigma \sigma, \sigma f_0, f_0 f_0$ decays, the highly smaller
direct CP-violating asymmetries are obtained for the corresponding $B_s^0$ decays in the
PQCD approach. As seen from the related decay amplitudes in Table~\ref{tab:DecayAmps-PS},
the fact is that, relative to the suppressed(enhanced) CKM matrix element $|V_{td}|(|V_{ud}|)$
in the $B_d^0$ decays, the enhanced(suppressed) one $|V_{ts}|(|V_{us}|)$ in the $B_s^0$ decays
contributes to the penguin(tree) amplitudes remarkably, which eventually weakened the interferences
between the tree and penguin amplitudes. Nevertheless, a bit large direct CP violation
$A_{\rm CP}^{\rm dir}(B_s^0 \to \sigma f_0) \sim -10\%$, associated with the large branching ratio
${\cal B}(B_s^0 \to \sigma f_0) \sim 10^{-4}$, could be tested at the LHCb and/or Belle-II
experiments in the (near) future.

\begin{figure}[!!htb]
\begin{center}
\hspace{-1 cm}
\includegraphics[scale=0.6]{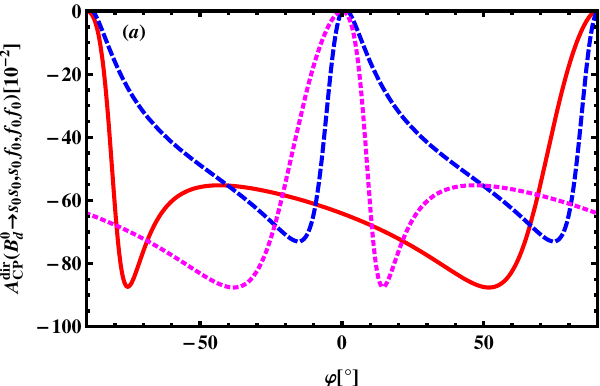}\hspace{1.2cm}
\includegraphics[scale=0.6]{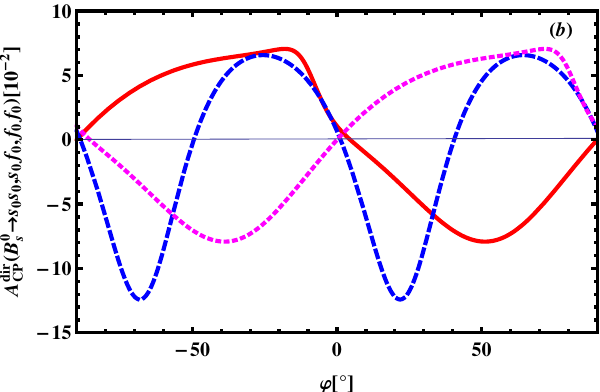}
\caption{(Color online) Dependence of the direct
CP-violating asymmetries
$A_{\rm CP}^{\rm dir}(B_{d,s}^0 \to \sigma \sigma,
\sigma f_0, f_0 f_0)$ on
$\varphi$ in the PQCD approach,
in which the red-solid line, the blue-dashed line, and the
magenta-dotted line correspond to
final states $\sigma \sigma$,
$\sigma f_0$, and $f_0 f_0$, respectively. }
\label{fig:fig3}
\end{center}
\end{figure}

Similarly, we also plot the variation of the direct CP-violating
asymmetries $A_{\rm CP}^{\rm dir}$ of the neutral $B$-meson decays
into $\sigma \sigma$, $\sigma f_0$, and $f_0 f_0$ with the mixing
angle $\varphi \in [-90^\circ, 90^\circ]$ in Fig.~\ref{fig:fig3}.
It is noted that, for the
$B_d^0$-meson decays, their direct CP violations
shown in Fig.~\ref{fig:fig3}(a) slightly change
from near $-40\% \sim -70\%$ at $\varphi \sim + 25^\circ$ to
about $-60\% \sim -70\%$ at $\varphi \sim - 25^\circ$; however,
for the $B_s^0$-meson decays, their direct CP violations
presented in Fig.~\ref{fig:fig3}(b) vary
dramatically with a total sign-changed, specifically,
$A_{\rm CP}^{\rm dir}(B_s^0 \to \sigma \sigma)$ from $-4\%$ to
$+ 7\%$, $A_{\rm CP}^{\rm dir}(B_s^0 \to \sigma \sigma)$ from $-12\%$ to
$+ 7\%$, and $A_{\rm CP}^{\rm dir}(B_s^0 \to \sigma \sigma)$ from $+ 5\%$ to
$- 6\%$, respectively. Therefore, we here also present the CP-violating
asymmetries for the $B_{d,s}^0 \to \sigma \sigma, \sigma f_0, f_0 f_0$ decays
at $\varphi \sim -25^\circ$ in the PQCD approach explicitly as follows:

\begin{itemize}
\item For the $B_d^0 \to \sigma \sigma, \sigma f_0, f_0 f_0$ decays,
\beq
A_{\rm CP}^{\rm dir}(B_d^0 \to \sigma \sigma) &=&
-57.38^{+6.91}_{-7.91}
\times 10^{-2}\;,
\qquad
\hspace{0.36cm}
A_{\rm CP}^{\rm mix}(B_d^0 \to \sigma \sigma) =
-78.41^{+4.83}_{-3.85}
\times 10^{-2}\;, \\
A_{\rm CP}^{\rm dir}(B_d^0 \to \sigma f_0) &=&
-66.68^{+9.15}_{-10.08}
\times 10^{-2}\;,
\qquad
\hspace{0.12cm}
A_{\rm CP}^{\rm mix}(B_d^0 \to \sigma f_0) =
-41.18^{+12.96}_{-10.58}
\times 10^{-2}\;,
\\
A_{\rm CP}^{\rm dir}(B_d^0 \to f_0 f_0) &=&
-71.77^{+13.74}_{-11.31}
\times 10^{-2}\;,
\qquad
A_{\rm CP}^{\rm mix}(B_d^0 \to f_0 f_0) =\hspace{0.30cm}
67.40^{+8.22}_{-10.62}
\times 10^{-2}\;.
\eeq

\item For the $B_s^0 \to \sigma \sigma, \sigma f_0, f_0 f_0$ decays,
\beq
A_{\rm CP}^{\rm dir}(B_s^0 \to \sigma \sigma) &=&\hspace{0.26cm}
6.83^{+0.56}_{-0.47}
\times 10^{-2}\;,
\qquad
\hspace{0.22cm}
A_{\rm CP}^{\rm mix}(B_s^0 \to \sigma \sigma) =\hspace{0.26cm}
8.69^{+1.17}_{-1.25}
\times 10^{-2}\;,
\\
A_{\rm CP}^{\rm dir}(B_s^0 \to \sigma f_0) &=&\hspace{0.26cm}
6.58^{+0.93}_{-0.88}
\times 10^{-2}\;,
\qquad
\hspace{0.12cm}
A_{\rm CP}^{\rm mix}(B_s^0 \to \sigma f_0) =\hspace{0.26cm}
7.04^{+1.55}_{-1.66}
\times 10^{-2}\;,
\\
A_{\rm CP}^{\rm dir}(B_s^0 \to f_0 f_0) &=&
-6.45^{+1.46}_{-1.51}
\times 10^{-2}\;,
\qquad
A_{\rm CP}^{\rm mix}(B_s^0 \to f_0 f_0) =\hspace{0.26cm}
2.24^{+0.75}_{-0.69}
\times 10^{-2}\;.
\eeq
and
\beq
A_{\rm CP}^{\rm \Delta\Gamma_s}(B_s^0 \to \sigma \sigma) &=&
99.39^{+0.12}_{-0.14}
\times 10^{-2}\;,
\qquad
A_{\rm CP}^{\rm \Delta\Gamma_s}(B_s^0 \to \sigma f_0) =
99.53^{+0.12}_{-0.10}
\times 10^{-2}\;,
\\
A_{\rm CP}^{\rm \Delta\Gamma_s}(B_s^0 \to f_0 f_0) &=&
99.77^{+0.09}_{-0.12}
\times 10^{-2}\;.
\eeq
\end{itemize}
By combining all the numerical results on the CP-averaged branching ratios
and the CP violations for the neutral $B$-meson decays into $\sigma \sigma$,
$\sigma f_0$, and $f_0 f_0$ at both of $\varphi \sim +25^\circ$ and $\varphi
\sim -25^\circ$ in the PQCD approach, it is expected that the near future
experiments could find some useful information on the magnitude and/or the sign
of the mixing angle $\varphi$, especially in the $B_s^0 \to \sigma \sigma$
and $\sigma f_0$ channels.

\bigskip
\section{Conclusions and Summary} \label{sec:summary}

In this paper, we have investigated the $B_{d,s}^0 \to \sigma \sigma, \sigma f_0,$
and $f_0 f_0$ decays through calculating the observables such as the CP-averaged
branching ratios and the CP-violating asymmetries within the framework of PQCD
approach. Due to the undetermined inner structure of the light scalars below 1~GeV in the
hadron sector, we made this PQCD analysis by considering the $\sigma$ and $f_0$
as the conventional two-quark-structure mesons in the product of $B$ meson decays.
With the referenced value $\sim \pm 25^\circ$ of the mixing
angle $\varphi$ in the quark-flavor basis, the numerical results in the
PQCD formalism show that all the six decay channels
of neutral $B$-meson decays into $\sigma \sigma, \sigma f_0$, and $f_0 f_0$ have
large decay rates and are expected to be confronted with the related experiments
through the four-body modes with $\sigma/f_0 \to \pi^+ \pi^-$ in the (near) future.
Of course, if the $f_0 \to K^+ K^-$ could be identified clearly from the
tail of the $\phi \to K^+ K^-$, then some of the four-body modes such as
$B_{d,s}^0 \to \sigma(\to \pi^+ \pi^-) f_0(\to K^+ K^-)$ and
$B_s^0 \to f_0(\to \pi^+ \pi^-/ K^+ K^-) f_0(\to K^+ K^-)$ could also
be examined in the future experiments.
It is worth mentioning that, different from those in the $B \to PP, PV, VV$ decays,
the non-factorizable emission diagrams of
the $B_{d,s}^0 \to \sigma \sigma, \sigma f_0, f_0 f_0$
decays in this work give large contributions because of the anti-symmetric
behavior of the leading-twist distribution amplitude of the scalar mesons.
The effective constraints from experiments and/or the reliable calculations
from Lattice QCD are very important for studying the light scalars in the
heavy meson decays.
Relative to the $B_s^0$ decays with suppressed tree amplitudes, the significant
interferences between both of the large
tree and penguin amplitudes in the $B_d^0$ decays contribute to the large
direct CP violations. It is expected that these related PQCD analyses
could provide useful information to constrain both magnitude and sign
of the mixing angle $\varphi$ between the $\sigma$ and $f_0$ with the
help of the future precise measurements. Honestly speaking, the determination
of the mixing angle $\varphi$ with its magnitude and sign indeed rely on the
sound constraints on the light-cone distribution amplitudes of scalar flavor
states $f_n$ and $f_s$ at both of theoretical and experimental aspects.
Of course, the possible final state interactions
or re-scattering effects, though existed as they should be, have to be left
for future studies elsewhere.


\begin{acknowledgments}

G.L. and H.N. thank Z.J. for helpful discussions.
This work is supported in part by the National Natural Science
Foundation of China under Grants  Nos.~11765012 and 11205072,
and by the Research Fund of Jiangsu Normal University under Grant No.~HB2016004.
G.L. and H.N. are supported by the
Undergraduate Research $\&$ Practice Innovation Program
of Jiangsu Province(No.~201910320018Z)

\end{acknowledgments}


\end{document}